\documentclass[journal]{IEEEtran}
\usepackage[T1]{fontenc}
\usepackage{ifpdf}
\usepackage{cite}
\usepackage[pdftex]{graphicx}
\usepackage{amsmath}
\usepackage{amssymb}
\usepackage{bm}
\usepackage{array}
\usepackage{fixltx2e}
\usepackage{stfloats}

\usepackage{balance} 

\usepackage{mathtools} 

\usepackage{diagbox}

\usepackage{makecell}

\usepackage{threeparttable}

\usepackage{color,soul}
\soulregister\cite7

\usepackage{algorithm}
\usepackage{algpseudocode}

\usepackage{setspace}

\usepackage{amsthm}

\newtheorem{theorem}{Theorem}
\newtheorem{proposition}{Proposition}
 
\newtheorem{lemma}{Lemma}
\newtheorem{assumption}{Assumption}

\newtheorem{remark}{Remark}

\makeatletter
\newcommand{\doublewidetilde}[1]{{%
  \mathpalette\double@widetilde{#1}%
}}
\newcommand{\double@widetilde}[2]{%
  \sbox\z@{$\m@th#1\widetilde{#2}$}%
  \ht\z@=.85\ht\z@
  \widetilde{\box\z@}%
}
\makeatother

\usepackage{bbm}

\usepackage{subcaption}

\usepackage[hidelinks]{hyperref}

\usepackage{bookmark}

\begin{document}

\title{\LARGE Robust IRS-Element Activation for Energy Efficiency Optimization in IRS-Assisted Communication Systems With Imperfect CSI}

\author{Christos N. Efrem and Ioannis Krikidis, \IEEEmembership{Fellow, IEEE} 
\thanks{This work received funding from the European Research Council (ERC) under the European Union's Horizon 2020 research and innovation programme (Grant agreement No. 819819) and the European Union's Horizon Europe programme (ERC, Grant agreement No. 101112697). It has also received funding from the European Union Recovery and Resilience Facility of the NextGenerationEU instrument, through the Research and Innovation Foundation under the project RISE (Grand Agreement: DUAL USE/0922/0031). Part of this work was presented at the IEEE International Conference on Communications (ICC), 2023 \cite{Efrem2023_ICC}.

The authors are with the Department of Electrical and Computer Engineering, University of Cyprus, 1678 Nicosia, Cyprus (e-mail: \{efrem.christos, krikidis\}@ucy.ac.cy). 

This article has been accepted for publication in \emph{IEEE Transactions on Wireless Communications}, May 2024. Copyright \copyright 2024 IEEE. Personal use is permitted, but republication/redistribution requires IEEE permission. 
}}


\maketitle

\begin{abstract}
In this paper, we study an intelligent reflecting surface (IRS)-aided communication system with single-antenna transmitter and receiver, under imperfect channel state information (CSI). More specifically, we deal with the robust selection of binary (on/off) states of the IRS elements in order to maximize the worst-case energy efficiency (EE), given a bounded CSI uncertainty, while satisfying a minimum signal-to-noise ratio (SNR). In addition, we consider not only continuous but also discrete IRS phase shifts. First, we derive closed-form expressions of the worst-case SNRs, and then formulate the robust (discrete) optimization problems for each case. In the case of continuous phase shifts, we design a dynamic programming (DP) algorithm that is theoretically guaranteed to achieve the global maximum with polynomial complexity $O(L\,{\log L})$, where $L$ is the number of IRS elements. In the case of discrete phase shifts, we develop a convex-relaxation-based method (CRBM) to obtain a feasible (sub-optimal) solution in polynomial time $O(L^{3.5})$, with a posteriori performance guarantee. Furthermore, numerical simulations provide useful insights and confirm the theoretical results. In particular, the proposed algorithms are several orders of magnitude faster than the exhaustive search when $L$ is large, thus being highly scalable and suitable for practical applications. Moreover, both algorithms outperform a baseline scheme, namely, the activation of all IRS elements.

\end{abstract}


\section{Introduction}

\subsection{Motivation and State-of-the-Art}

Intelligent reflecting surfaces (IRSs) have emerged as a promising technology for the dynamic configuration of electromagnetic waves \cite{Liaskos2018,DiRenzo2019,Wu2020}. In particular, IRSs can control the amplitude and/or the phase of the incident radio waves by using tunable electronic circuits, such as positive intrinsic negative (PIN) or varactor diodes. In the former case this is achieved by appropriately configuring the state of the PIN diodes, while in the latter case by changing the bias voltage of the varactors. A key characteristic of passive IRSs is low energy consumption. Energy efficiency (EE) optimization has been an attractive area of research in wireless networks \cite{Efrem2019a,Efrem2019b,Efrem2020}, and more recently in IRS-aided communication systems \cite{Huang2019}. Moreover, there has been a growing interest in active/passive-IRS beamforming to optimize the achievable sum rate \cite{Zhao2021}, signal-to-noise ratio (SNR) \cite{Long2021,Zhang2022}, and transmit power \cite{Wu2019,Zhou2020}. 

In addition, the joint optimization of IRS location and size (i.e., the number of IRS elements) to minimize the outage probability has been studied in \cite{Efrem2022,Efrem2023}. By considering channel estimation and feedback, the optimum number of IRS elements maximizing the spectral efficiency (SE), EE, and SE-EE trade-off has been investigated in \cite{Zappone2021}. Also, the author in \cite{Li2022} has examined the minimum number of IRS elements satisfying minimum-SE/EE requirements. A low-cost scheme for optimizing only the on/off states of IRS elements, while fixing their phase shifts, has been given in \cite{Khaleel2022}.

Furthermore, realistic power consumption models for different types of IRSs with measurement validation have been proposed in \cite{Wang2023}. EE maximization problems, considering the aforementioned practical model for PIN-diode-based IRS, have been studied in \cite{Li2023, Xu2023} to optimize the base-station beamforming and IRS phase-shift matrices. Moreover, the authors in \cite{Yang2022} and \cite{Al-Nahhas2021} have conducted performance analyses of IRS-assisted single-input single-output (SISO) and multiple-input single-output (MISO) systems, respectively, under imperfect CSI. The joint transmit beamforming and IRS phase-shift design has been examined in \cite{Gao2021} to minimize the total power under a bounded channel-estimation-error model, and in \cite{Gao2023} to maximize the sum rate assuming a practical phase-shift model and statistical CSI error. It is worth noting that the authors in \cite{Al-Nahhas2021, Gao2021, Gao2023} have derived exact/approximate expressions for the signal-to-interference-plus-noise ratio (SINR) or achievable rate in multi-antenna scenarios, regardless of optimizing the on/off states of IRS elements.

\subsection{Key Contributions}

This paper deals with the robust optimization of EE under imperfect channel state information (CSI), assuming either \emph{continuous} or \emph{discrete} IRS phase shifts. The main contributions are summarized as follows: 

\begin{itemize}

\item First of all, we derive a \emph{low-complexity formula for the computation of discrete phase shifts}, based on the closest point rule (see Theorem \ref{theorem:discrete_phase-shifts_closed-form}). Specifically, the complexity of this method is $O(L b)$, where $L$ and $b$ are the number of IRS elements and quantization bits, respectively. In fact, this formula achieves a significant complexity reduction (i.e., an \emph{exponential speed-up} with respect to $b$) compared to the one-by-one inspection of the decision regions for each IRS element, which requires $O(L {2^b})$ time. It also turns out that the proposed formula is \emph{asymptotically optimal} (see Remark \ref{remark:Fastest_possible_formula}).

\item In addition, we provide \emph{closed-form expressions of the worst-case SNRs under CSI uncertainty} (see Theorems \ref{theorem:Worst-case_SNR_discrete} and \ref{theorem:Worst-case_SNR_continuous}), which are then used to formulate the robust optimization problems. In particular, we aim to maximize the worst-case EE, subject to a minimum-SNR constraint, by tuning the on/off states of IRS elements. The IRS phase shifts are adjusted dynamically using the estimated channels. On the other hand, existing works dealing with the activation of IRS elements assume perfect CSI or only phase errors \cite{Zappone2021,Khaleel2022}, fixed IRS phase shifts \cite{Khaleel2022}, optimize different performance metrics (other than EE) \cite{Zhao2021,Li2022,Khaleel2022}, and do not consider a minimum-SNR constraint \cite{Zappone2021}. Note also that other works \cite{Li2023, Xu2023, Al-Nahhas2021, Yang2022, Gao2021, Gao2023} do not study the optimization of the on/off states of IRS elements at all. 

\item In the case of continuous phase shifts, we develop a \emph{polynomial-time dynamic programming (DP) algorithm with global optimization guarantee} (see Theorem \ref{theorem:DP_algorithm}). To the best of our knowledge, this is the first work that achieves global optimization of EE (with polynomial complexity) in IRS-aided communication systems by taking into account imperfect CSI knowledge. 

\item In the case of discrete phase shifts, we first formulate a concave-affine fractional relaxation problem, and then apply the Charnes-Cooper transform to generate an equivalent convex problem. Subsequently, we design a \emph{polynomial-time convex-relaxation-based method (CRBM) that computes a sub-optimal solution whose objective value is within a maximum distance from the global optimum} (see Theorem \ref{theorem:CRBM}).

\item Finally, numerical results show that: i) the DP algorithm has identical performance with the exhaustive search, ii) CRBM attains near-optimal performance, and iii) the proposed algorithms perform better than a conventional scheme, while achieving remarkable trade-offs between performance and complexity (since their running time scales mildly with the number of IRS elements).     

\end{itemize}

\subsection{Outline}

The rest of this paper is organized as follows. Section \ref{section:System_Model} describes the system model and Section \ref{section:Problem_Formulation} formulates the robust optimization problems under CSI uncertainty. Afterwards, Sections \ref{section:Dynamic_Programming} and \ref{section:Approximation_Algorithm} develop and analyze the proposed algorithms. Moreover, Section \ref{section:Numerical_Results} presents some numerical results, while Section \ref{section:Conclusion} concludes the paper. Finally, Appendices include lengthy mathematical proofs.

\subsection{Mathematical Notation}

Italic lowercase letters denote scalars, boldface lowercase (or uppercase) letters stand for column vectors (respectively, matrices), and calligraphic uppercase letters represent sets. $\left| z \right| \geq 0$ and $\operatorname{Arg}(z) \in [0,2\pi)$ denote the complex magnitude (or modulus) and the principal argument of a complex number $z$, respectively, while $j = \sqrt{-1}$ is the imaginary unit. Moreover, ${{\mathbf{0}}_N}$/$\mathbf{1}_N$ is the $N$-dimensional zero/all-ones vector, $\left[\cdot \right]^\top$ stands for matrix transpose, and ${\operatorname{diag}}(\mathbf{x})$ denotes a diagonal matrix with the elements of $\mathbf{x}$ on its main diagonal. ${\lVert \mathbf{z} \rVert}_2 = \sqrt{ \sum_{n=1}^{N} {|z_n|^2} }$ is the Euclidean norm of a complex vector $\mathbf{z} \in \mathbb{C}^N$. The symbols $\triangleq$ and $\sim$ mean ``by definition equal to'' and ``distributed as'', respectively. Furthermore, $\log( \cdot )$ is the natural logarithm (i.e., with base $e$), and $\binom{n}{m} = \tfrac{n!}{m!(n - m)!}$ is the binomial coefficient. $O(\cdot)$, $\Omega(\cdot)$, and $\Theta(\cdot)$ are respectively the big-omicron, big-omega, and big-theta standard asymptotic notation. 

In addition, for any $y \in \mathbb{R}$ and any $a > 0$, the modulo operation is defined as $y \bmod a = y - a \lfloor y/a \rfloor$, where $\lfloor \cdot \rfloor$ is the floor function. Note that $y \bmod {a} \in [0,a)$, because $r - 1 < \lfloor r \rfloor \leq r$, for all $r \in \mathbb{R}$. Specifically, for any integer $m$ and any positive integer $n$, it holds that $m \bmod n \in \{0,\dots,n-1\}$. Finally, $\land$ and $\lor$ stand for logical conjunction (AND) and logical disjunction (OR), respectively.

\section{System Model} \label{section:System_Model}

\subsection{Signal Transmission Model}

Consider a wireless system consisting of a single-antenna transmitter (Tx) and a single-antenna receiver (Rx) that communicate through a passive IRS with $L$ elements.\footnote{The scenario of single-antenna Tx and Rx is fundamental and serves as a benchmark (i.e., reference point) for comparison with more complex setups. The extension of our results to multi-user multi-antenna systems is not straightforward and deserves further investigation. Nevertheless, the proposed algorithms take into consideration the impact of imperfect CSI not only for continuous but also for discrete phase shifts to optimize the on/off states of IRS elements.}\textsuperscript{,}\footnote{An IRS is called ``passive'' if all its elements are passive (i.e., their induced amplitude is less than or equal to 1). On the other hand, it is called ``active'' if some of its elements are active (i.e., their amplitude is greater than 1).} Without loss of generality, the set of all IRS elements is denoted by $\mathcal{L} = \{1,\dots,L\}$. Let $h_0 \in \mathbb{C}$ be the channel coefficient of the Tx-Rx direct link, and ${{\mathbf{u}}} = {[{u_1}, \ldots ,{u_L}]^ \top } \in \mathbb{C}^L$, ${{\mathbf{v}}} = {[{v_1}, \ldots ,{v_L}]^ \top \in \mathbb{C}^L}$ be the channel-coefficient vectors of Tx-IRS and IRS-Rx links, respectively. All channel coefficients are constant within a given time-slot (flat-fading channels). 

Moreover, each IRS element can be either on (i.e., activated) or off (i.e., deactivated). In particular, we say that an IRS element is \emph{``on''} when operating in standard reflection mode (\emph{reflecting element}), whereas it is \emph{``off''} when functioning in absorption mode (\emph{absorbing element}) \cite{Zhao2021,Khaleel2022}. For this purpose, we introduce a binary vector ${{\mathbf{x}}} = {[{x_1}, \ldots ,{x_L}]^\top \in \{0,1\}^L}$, so that $x_\ell = 1$ if and only if the $\ell^\text{th}$ IRS element is on; equivalently, $x_\ell = 0$ if and only if the $\ell^\text{th}$ IRS element is off. The IRS on/off-state-and-phase-shift diagonal matrix is given by ${{\mathbf{D}}} (\mathbf{x},\boldsymbol{\phi}) = {\operatorname{diag}}({x_1 \beta_1 e^{j{\phi_1}}}, \ldots ,{x_L \beta_L e^{j{\phi _{L}}}}) \in \mathbb{C}^{L \times L}$, where $\beta_\ell \in [0,1]$ is the amplitude attenuation and $\phi_\ell \in [0,2\pi)$ is the phase shift of the $\ell^{\text{th}}$ IRS element (with ${\boldsymbol{\phi}} = [{\phi_1}, \ldots ,{\phi_L}]^\top \in [0,2\pi)^L$). In this work, we assume that the amplitude coefficients $\{ \beta_\ell \}_{\ell \in \mathcal{L}}$ are fixed.   

We also define the cascaded channel coefficient corresponding to the $\ell^\text{th}$ IRS element, i.e., $h_\ell = {\beta_\ell} {u_\ell} {v_\ell} \in \mathbb{C}$, for all $\ell \in \mathcal{L}$. The complex channel coefficients (direct and cascaded) are expressed in polar coordinates, i.e., ${h_\ell} = {\alpha_\ell} {e^{j\theta_\ell}}$, where $\alpha_\ell = | h_\ell | \geq 0$ and $\theta_\ell = \operatorname{Arg}(h_\ell) \in [0,2\pi)$, for all $\ell \in \mathcal{L}_0 = \{0,\dots,L\}$. For convenience, the direct and cascaded channel coefficients are grouped in vector form as $\mathbf{h} = [h_0,\dots,h_L]^\top \in \mathbb{C}^{L+1}$.

The received signal is a superposition of the transmitted signal through the direct and cascaded channels, i.e.,
\begin{equation}
\begin{split}
r & = \sqrt{p} \left(h_0 + \mathbf{u}^\top {\mathbf{D}}(\mathbf{x},\boldsymbol{\phi}) \, \mathbf{v} \right) s + w  \\
& = \sqrt{p} \left( h_0 + \sum_{\ell \in \mathcal{L}} {x_\ell h_\ell  e^{j\phi_\ell} } \right)s + w   ,
\end{split}  
\end{equation}
where $p > 0$ is the (fixed) transmit power, $s \in \mathbb{C}$ is the transmitted symbol (with $\mathbb{E}(|s|^2)=1$), and $w \sim \mathcal{CN}(0,\sigma_w^2)$ is the additive white Gaussian noise (AWGN) at the receiver (with ${\sigma_w^2}>0$ being the noise power). Therefore, the SNR at the receiver is given by 
\begin{equation} \label{equation:SNR_initial}
\gamma = \overline{\gamma} \left|  h_0 + \sum_{\ell \in \mathcal{L}} {x_\ell h_\ell  e^{j\phi_\ell} }  \right| ^2 ,
\end{equation}
where $\overline{\gamma} = p/{\sigma_w^2} > 0$.

\subsection{Imperfect CSI Model}

Regarding the CSI acquisition, we assume that the receiver estimates the direct and cascaded channel coefficients, and then feeds the vector-quantized CSI back to the transmitter. As a result, the obtained CSI is subject to quantization errors.

Herein, we adopt an additive CSI-error model, i.e., 
\begin{equation} \label{equation:CSI-error_model}
\mathbf{h} = \widehat{\mathbf{h}} + \widetilde{\mathbf{h}}  ,
\end{equation}
where $\widehat{\mathbf{h}} = [\widehat{h}_0,\dots,\widehat{h}_L]^\top \in \mathbb{C}^{L+1}$ is the estimated channel vector (\emph{known} at the transmitter), while $\widetilde{\mathbf{h}} = [\widetilde{h}_0,\dots,\widetilde{h}_L]^\top \in \mathbb{C}^{L+1}$ is the \emph{unknown} channel estimation error. As before, we can express these complex coefficients in polar coordinates, i.e., ${\widehat{h}_\ell} = {\widehat{\alpha}_\ell} {e^{j\widehat{\theta}_\ell}}$ and ${\widetilde{h}_\ell} = {\widetilde{\alpha}_\ell} {e^{j\widetilde{\theta}_\ell}}$, for all $\ell \in \mathcal{L}_0$.

In order to avoid making any assumptions on the statistics of $\widetilde{\mathbf{h}}$, we consider a (deterministic) \emph{bounded CSI-error model}. More specifically, we assume that $\widetilde{\mathbf{h}}$ lies within a compact (closed and bounded) ball of uncertainty, i.e.,
\begin{equation} \label{equation:CSI_uncertainty}
{\lVert \widetilde{\mathbf{h}} \rVert}_2 \triangleq \sqrt{ \sum_{\ell \in \mathcal{L}_0} {\widetilde{\alpha}_\ell^2} } \leq \delta  ,
\end{equation} 
where $\delta \geq 0$ is the radius of the CSI-uncertainty region \emph{known} at the transmitter \cite{Zhou2020,Shenouda2007}.\footnote{Other works in the literature \cite{Zhou2020,Wang2014,Ntougias2022} assume a priori (known in advance) probability distribution for the estimation error, e.g., circularly symmetric complex Gaussian distribution, which results in a less conservative probabilistic constraint compared to the worst-case constraint in \eqref{equation:CSI_uncertainty}.} Note that by setting $\delta = 0$ we obtain the case of \emph{perfect CSI} (i.e., without channel estimation errors), since $\delta = 0$ implies $\widetilde{\mathbf{h}} = \mathbf{0}_{L+1}$ and therefore $\widehat{\mathbf{h}} = \mathbf{h}$.

Furthermore, we make the following assumption that will be useful throughout the paper. 

\begin{assumption} \label{assumption:CSI-uncertainty_radius}
The CSI-uncertainty radius is less than or equal to the minimum magnitude of estimated channel coefficients, i.e., $\delta \leq \widehat{\alpha}_{\min}$, where $\widehat{\alpha}_{\min} = \min_{\ell \in \mathcal{L}_0} \{ \widehat{\alpha}_\ell \}$. 
\end{assumption}

\begin{remark} 
Assumption \ref{assumption:CSI-uncertainty_radius} can be easily checked in practice, since $\{ \widehat{\alpha}_\ell \}_{\ell \in \mathcal{L}_0}$ belong to a finite (discrete) set of quantization levels that is known in advance.
\end{remark}

\subsection{IRS Phase-Shift Design}

Under the CSI-error model given by \eqref{equation:CSI-error_model}, the SNR in \eqref{equation:SNR_initial} can be written as 
\begin{equation} \label{equation:SNR_with_estimation_error}
\gamma (\mathbf{x},\widetilde{\mathbf{h}},\boldsymbol{\phi}) = \overline{\gamma} \left|  \widehat{h}_0 + \sum_{\ell \in \mathcal{L}} {x_\ell \widehat{h}_\ell e^{j\phi_\ell} } + \widetilde{h}_0 + \sum_{\ell \in \mathcal{L}} {x_\ell \widetilde{h}_\ell e^{j\phi_\ell} } \right| ^2 .
\end{equation}

\subsubsection{Continuous Phase Shifts}

If the number of bits used for controlling the IRS phase shifts is sufficiently large (infinite bit-resolution), then the quantization error becomes negligible and the phase shifts can be considered continuous. 

Since the channel estimation error $\widetilde{\mathbf{h}}$ is unknown, the IRS phase shifts are adjusted so as to maximize the ideal SNR without CSI error ($\widetilde{\mathbf{h}} = \mathbf{0}_{L+1}$), using only the estimated channel $\widehat{\mathbf{h}}$. Therefore, the optimal phase shifts are given by\footnote{Here, we use the modulo operation in order to ensure that $\phi_\ell^\star \in [0,2\pi)$.}
\begin{equation} 
\phi_\ell^{\star} =  (\widehat{\theta}_0 - \widehat{\theta}_\ell) \bmod {2\pi}, \;\; \forall \ell \in \mathcal{L} . 
\end{equation} 
Based on \eqref{equation:SNR_with_estimation_error}, the SNR achieved by the optimal (continuous) phase shifts is expressed as 
\begin{equation} \label{equation:SNR_continuous_phase_shifts}
\begin{split}
{\gamma}^\text{c} (\mathbf{x},\widetilde{\mathbf{h}}) & \triangleq \gamma (\mathbf{x},\widetilde{\mathbf{h}},{\boldsymbol{\phi}}^{\star}) \\
& = \overline{\gamma} \left| f_\text{c}(\mathbf{x}) e^{j\widehat{\theta}_0} + \widetilde{h}_0 + \sum_{\ell \in \mathcal{L}} {x_\ell \widetilde{h}_\ell e^{j{\phi}_\ell^{\star}}} \right| ^2  ,
\end{split}
\end{equation}
where  
\begin{equation} \label{equation:f_c}
f_\text{c}(\mathbf{x}) =  \widehat{\alpha}_0 + \sum_{\ell \in \mathcal{L}} {x_\ell \widehat{\alpha}_\ell}  .
\end{equation}

\subsubsection{Discrete Phase Shifts}

Consider a finite and nonzero number of quantization bits, $b$ (i.e., a $b$-bit phase-shift resolution). In this case, the number of quantization levels is $K=2^b$. Assuming \emph{uniform} quantization, the IRS phase shifts belong to the following discrete set 
\begin{equation}
{\Phi_K} = \{0, \omega, 2\omega, \dots, (K-1)\omega \} \subseteq [0,2\pi)  ,
\end{equation}
where $\omega = 2\pi / K$. The following mild assumption will be useful in the rest of the paper.

\begin{assumption} \label{assumption:minimum_quantization_bits}
The number of quantization bits is greater than or equal to two (i.e., $b \geq 2$), and therefore $K \geq 4$.   
\end{assumption}

Now, we define the following (non-overlapping/disjoint) \emph{decision regions}:
\begin{gather}
\mathcal{R}_0 = [0,\omega/2) \cup [2\pi - \omega/2, 2\pi) , \\
\mathcal{R}_k = [k\omega - \omega/2, k\omega + \omega/2) , \;\; \forall k \in \{1,\dots,K-1\} .
\end{gather}
In other words, the sets $\mathcal{R}_k$, $k \in \{0,\dots,K-1 \}$, form a partition of the set $[0,2\pi)$: $\mathcal{R}_k \neq \varnothing$ for all $k$, $\mathcal{R}_m \cap \mathcal{R}_n = \varnothing$ whenever $m \neq n$, and $\bigcup_{k=0}^{K-1} {\mathcal{R}_k} = [0,2\pi)$. The discrete phase shifts are chosen according to the \emph{closest point rule} (i.e., by rounding the optimal phase shift to the closest point in the discrete set $\Phi_K$) as follows 
\begin{equation} \label{equation:discrete_phase-shifts}
{\phi}^{\text{d}}_\ell = \lambda_\ell \, \omega , \;\; \forall \ell \in \mathcal{L} ,
\end{equation}
where $\lambda_\ell$ is the \emph{unique} integer (because the decision sets are disjoint) that satisfies $\lambda_\ell \in \{ 0,\dots,K-1 \}$ and $\phi_\ell^\star \in \mathcal{R}_{\lambda_\ell}$.\footnote{Note that \eqref{equation:discrete_phase-shifts} is a slightly different rule compared to that used so far in the literature: ${\phi}^{\text{d}}_\ell \in \mathop{\arg \min}_{\phi \in {\Phi_K}} |\phi - \phi_\ell^\star| ,\ \forall \ell \in \mathcal{L}$, which is referred to as the ``closest point projection'' \cite{Zhang2022}. The latter rule has two major disadvantages: i) the minimum distance can be attained at more than one discrete value, e.g., when $\phi_\ell^\star = k\omega + \omega/2$ for some $k \in \{ 0,1,\dots,K-1 \}$, and ii) if $\phi_\ell^\star \in (2\pi - \omega/2,2\pi)$, then ${\phi}^{\text{d}}_\ell = (K-1)\omega = 2\pi - \omega$ instead of $0$; this error is negligible when $K$ is relatively large, but becomes significant when $K$ is small because there is a higher probability that $\phi_\ell^\star \in (2\pi - \omega/2,2\pi)$. On the other hand, the proposed rule in \eqref{equation:discrete_phase-shifts} does not suffer from such weaknesses.}

\begin{theorem} \label{theorem:discrete_phase-shifts_closed-form}
Given that ${\boldsymbol{\phi}}^{\star} \in [0,2\pi)^L$, the closest point rule in \eqref{equation:discrete_phase-shifts} for selecting discrete phase shifts is equivalent to 
\begin{equation} \label{equation:discrete_phase-shifts_closed-form}
{\phi}^{\textnormal{d}}_\ell = [\operatorname{round}({\phi_\ell^\star} / \omega) \bmod K] \, \omega  , \;\; \forall \ell \in \mathcal{L} ,
\end{equation} 
where $\operatorname{round}(x) = \lfloor x + 1/2 \rfloor$. In addition, the closed-form expression \eqref{equation:discrete_phase-shifts_closed-form} requires $O(L \log K) = O(L b)$ arithmetic operations, thus achieving an exponential speed-up (with respect to $K$ or $b$) compared to the $O(L K) = O(L {2^b})$ complexity of \eqref{equation:discrete_phase-shifts} that results from the one-by-one inspection of the decision regions for each $\ell \in \mathcal{L}$. 
\end{theorem}

\begin{IEEEproof}
See Appendix \ref{appendix:discrete_phase-shifts_closed-form}.    
\end{IEEEproof}

\begin{remark}  \label{remark:Fastest_possible_formula}
Since any algorithm that outputs an integer in the set $\{ 0,1,\dots,K-1 \}$ has complexity $\Omega (\log K) = \Omega (b)$ in the worst case, the proposed formula given by \eqref{equation:discrete_phase-shifts_closed-form} is fastest/best-possible in terms of asymptotic complexity. 
\end{remark}

The phase-shift quantization error, between discrete and optimal phase shifts, is defined by 
\begin{equation}
\varepsilon_\ell = {\phi}^{\text{d}}_\ell - \phi_\ell^\star ,  \;\; \forall \ell \in \mathcal{L} .
\end{equation}
Notice that $\varepsilon_\ell \in (-\omega/2,\omega/2]$, thus $|\varepsilon_\ell| \in [0,\omega/2],\ \forall \ell \in \mathcal{L}$.

By virtue of \eqref{equation:SNR_with_estimation_error}, the SNR achieved by the discrete phase shifts is given by
\begin{equation} \label{equation:SNR_discrete_phase_shifts_initial}
\begin{split}
{\gamma}^\text{d} (\mathbf{x},\widetilde{\mathbf{h}}) & \triangleq \gamma (\mathbf{x},\widetilde{\mathbf{h}},{\boldsymbol{\phi}}^\text{d}) \\
&= \overline{\gamma} \Bigg| \left( \widehat{\alpha}_0 + \sum_{\ell \in \mathcal{L}} {x_\ell \widehat{\alpha}_\ell e^{j\varepsilon_\ell}} \right) e^{j\widehat{\theta}_0}  \\
&\quad + \widetilde{h}_0 + \sum_{\ell \in \mathcal{L}} {x_\ell \widetilde{h}_\ell e^{j{\phi}_\ell^{\text{d}}} } \Bigg| ^2  ,
\end{split}
\end{equation}
or, equivalently, 
\begin{equation} \label{equation:SNR_discrete_phase_shifts}
{\gamma}^\text{d} (\mathbf{x},\widetilde{\mathbf{h}}) = \overline{\gamma} \left| f_\text{d}(\mathbf{x})  e^{j(\vartheta(\mathbf{x})+\widehat{\theta}_0)} + \widetilde{h}_0 + \sum_{\ell \in \mathcal{L}} {x_\ell \widetilde{h}_\ell e^{j{\phi}_\ell^{\text{d}}} } \right| ^2  ,
\end{equation} 
where 
\begin{equation} \label{equation:f_d}
f_\text{d}(\mathbf{x}) = \left|  \widehat{\alpha}_0 + \sum_{\ell \in \mathcal{L}} {x_\ell \widehat{\alpha}_\ell e^{j\varepsilon_\ell}}  \right|  ,
\end{equation}
\begin{equation}
\vartheta(\mathbf{x}) = \operatorname{Arg}  \left(  \widehat{\alpha}_0 + \sum_{\ell \in \mathcal{L}} {x_\ell \widehat{\alpha}_\ell e^{j\varepsilon_\ell}}  \right) .
\end{equation}

\begin{proposition} \label{proposition:limits}
As the number of quantization bits tends to infinity, we have the following limits 
\begin{equation} \label{equation:limits_K_omega}
\lim_{b \to \infty} {K} = \infty  ,  \;\; \lim_{b \to \infty} {\omega} = 0  ,
\end{equation}
\begin{equation} \label{equation:limits_epsilon_phi} 
\lim_{b \to \infty} {\varepsilon_\ell} = 0 , \;\;  \lim_{b \to \infty} {{\phi}^{\textnormal{d}}_\ell} = \phi_\ell^\star, \;\; \forall \ell \in \mathcal{L} ,
\end{equation}
\begin{equation} \label{equation:limits_gamma}
\lim_{b \to \infty} {{\gamma}^\textnormal{d} (\mathbf{x},\widetilde{\mathbf{h}})} = {\gamma}^\textnormal{c} (\mathbf{x},\widetilde{\mathbf{h}}), \;\; \forall \mathbf{x} \in \{0,1\}^L , \forall \widetilde{\mathbf{h}} \in \mathbb{C}^{L+1}  , 
\end{equation}
\begin{equation} \label{equation:limits_f}
\lim_{b \to \infty} {f_\textnormal{d}(\mathbf{x})} = f_\textnormal{c}(\mathbf{x}), \;\; \forall \mathbf{x} \in \{0,1\}^L . 
\end{equation}
Furthermore, provided that $\widehat{\alpha}_0 > 0$, if $\overline{\vartheta} (\mathbf{x})$ is an accumulation point of the sequence $\{\vartheta(\mathbf{x};b)\}_{b \geq 1}$, then $\overline{\vartheta} (\mathbf{x}) \in \{0,2\pi\}$, for all $\mathbf{x} \in \{0,1\}^L$. Here, with a slight abuse of notation, we write $\vartheta(\mathbf{x};b)$ in place of $\vartheta(\mathbf{x})$ to denote the dependence on $b$. 
\end{proposition}

\begin{IEEEproof}
The limits in \eqref{equation:limits_K_omega} are straightforward, based on the definitions of $K$ and $\omega$. In addition, $\lim_{b \to \infty} {\varepsilon_\ell} = 0$ because $-\omega/2 < \varepsilon_\ell \leq \omega/2$, while the definition of $\varepsilon_\ell$ implies that $\lim_{b \to \infty} {{\phi}^{\textnormal{d}}_\ell} = \phi_\ell^\star$ (note that $\phi_\ell^\star$ is independent of $b$). Hence, \eqref{equation:limits_epsilon_phi} has been proved. 

By taking the limit as $b \to \infty$ in \eqref{equation:SNR_discrete_phase_shifts_initial}, we deduce that $\lim_{b \to \infty} {{\gamma}^\textnormal{d} (\mathbf{x},\widetilde{\mathbf{h}})} = {\gamma}^\textnormal{c} (\mathbf{x},\widetilde{\mathbf{h}})$, due to  the continuity of the complex magnitude and because of  \eqref{equation:limits_epsilon_phi}; see also \eqref{equation:SNR_continuous_phase_shifts}. Moreover, by taking the limit in \eqref{equation:f_d}, we obtain $\lim_{b \to \infty} {f_\textnormal{d}(\mathbf{x})} = f_\textnormal{c}(\mathbf{x})$ for similar reasons. 

Now, suppose that $\widehat{\alpha}_0 > 0$. Then, $f_\textnormal{c}(\mathbf{x}) > 0$, for all $\mathbf{x} \in \{0,1\}^L$. Let $\overline{\vartheta} (\mathbf{x})$ be an accumulation point of the sequence $\{\vartheta(\mathbf{x};b)\}_{b \geq 1}$, i.e., $\lim_{b \to \infty, \, b \in \mathcal{B}} {\vartheta(\mathbf{x};b)} = \overline{\vartheta} (\mathbf{x})$ for some (countably) infinite set $\mathcal{B}$ of distinct natural numbers. Note that $\overline{\vartheta} (\mathbf{x}) \in [0,2\pi]$, because $\vartheta(\mathbf{x};b) = \vartheta(\mathbf{x}) \in [0,2\pi)$. If we  take the limit as $b \to \infty$, with $b \in \mathcal{B}$, in $\widehat{\alpha}_0 + \sum_{\ell \in \mathcal{L}} {x_\ell \widehat{\alpha}_\ell e^{j\varepsilon_\ell}} = f_\textnormal{d}(\mathbf{x}) e^{j \vartheta(\mathbf{x})} = f_\textnormal{d}(\mathbf{x}) \cos(\vartheta(\mathbf{x})) + j f_\textnormal{d}(\mathbf{x}) \sin(\vartheta(\mathbf{x}))$ and exploit the continuity of sine and cosine, we have that
\begin{equation}
f_\textnormal{c} (\mathbf{x}) = f_\textnormal{c}(\mathbf{x}) \cos(\overline{\vartheta} (\mathbf{x})) + j f_\textnormal{c}(\mathbf{x}) \sin(\overline{\vartheta} (\mathbf{x})) .
\end{equation}
By equating the real and imaginary parts of both sides and since $f_\textnormal{c}(\mathbf{x}) > 0$, we obtain $(\cos(\overline{\vartheta} (\mathbf{x})) = 1) \land (\sin(\overline{\vartheta} (\mathbf{x})) = 0)$, thus yielding $(\overline{\vartheta} (\mathbf{x}) = 0) \lor (\overline{\vartheta} (\mathbf{x}) = 2\pi)$. 
\end{IEEEproof}

Next, by using \emph{Euler's formula}: $e^{j\varepsilon_\ell} = \cos(\varepsilon_\ell) + j\sin(\varepsilon_\ell)$, we can write 
\begin{equation} \label{equation:f_d_sqrt}
f_\text{d}(\mathbf{x}) = \sqrt{ \Delta_{\Re}^2(\mathbf{x}) + \Delta_{\Im}^2(\mathbf{x}) }  ,
\end{equation}
where $\Delta_{\Re}(\mathbf{x}) = \widehat{\alpha}_0 + \sum_{\ell \in \mathcal{L}} {x_\ell \widehat{\alpha}_\ell \cos (\varepsilon_\ell)}$ and $\Delta_{\Im}(\mathbf{x}) = \sum_{\ell \in \mathcal{L}} {x_\ell \widehat{\alpha}_\ell \sin (\varepsilon_\ell)}$. This expression of $f_\textnormal{d}(\mathbf{x})$ yields the following proposition.

\begin{proposition}
The function $f_\textnormal{d}(\mathbf{x})$ can be expanded in the following form  
\begin{equation} \label{equation:f_d_sqrt_expanded}
f_\textnormal{d}(\mathbf{x}) = \sqrt{ \widehat{\alpha}_0^2 + \sum_{\ell \in \mathcal{L}}{{\zeta_\ell} {x_\ell}} +  \mathop{\sum\sum}_{1 \leq n < m \leq L} {{\mu_{n m}} {x_n} {x_m}} }  ,
\end{equation}
where $\zeta_\ell = \widehat{\alpha}_\ell^2 + 2 \widehat{\alpha}_0 \widehat{\alpha}_\ell \cos(\varepsilon_\ell)$, for all $\ell \in \mathcal{L}$, and $\mu_{n m} = 2 \widehat{\alpha}_n \widehat{\alpha}_m \cos(\varepsilon_n - \varepsilon_m)$ whenever $1 \leq n < m \leq L$.\footnote{Note that $\mathop{\sum\sum}_{1 \leq n < m \leq L} \{\cdot\} = \sum_{n=1}^{L-1} {\sum_{m=n+1}^L} \{\cdot\} =  \sum_{m=2}^L {\sum_{n=1}^{m-1}} \{\cdot\}$.} In addition, $f_\textnormal{d}(\mathbf{x})$ can be equivalently written as
\begin{equation} \label{equation:f_d_sqrt_expanded_min}
f_\textnormal{d}(\mathbf{x}) = \sqrt{ \widehat{\alpha}_0^2 + \sum_{\ell \in \mathcal{L}}{{\zeta_\ell} {x_\ell}} + \mathop{\sum\sum}_{1 \leq n < m \leq L} {{\mu_{n m}} {\min({x_n},{x_m})}} }  .
\end{equation}
\end{proposition}

\begin{IEEEproof}
By leveraging the following \emph{square-of-sum identity} of $N$ real numbers $\{a_i\}_{i=1}^N$
\begin{equation}
\left( \sum_{i=1}^N {a_i} \right)^2 = \sum_{i=1}^N {a_i^2} + 2 \sum_{n=1}^{N-1} {\sum_{m=n+1}^N} {a_n a_m} ,
\end{equation}
we can express $\Delta_{\Re}^2(\mathbf{x})$ and $\Delta_{\Im}^2(\mathbf{x})$ as follows
\begin{equation}
\begin{split}
\Delta_{\Re}^2(\mathbf{x}) = & \; \widehat{\alpha}_0^2 + 2 \widehat{\alpha}_0 \sum_{\ell \in \mathcal{L}} {x_\ell \widehat{\alpha}_\ell \cos (\varepsilon_\ell)} + \left( \sum_{\ell \in \mathcal{L}} {x_\ell \widehat{\alpha}_\ell \cos (\varepsilon_\ell)} \right)^2   \\
= & \; \widehat{\alpha}_0^2 + 2 \widehat{\alpha}_0 \sum_{\ell \in \mathcal{L}} {x_\ell \widehat{\alpha}_\ell \cos (\varepsilon_\ell)} + \sum_{\ell \in \mathcal{L}} {x_\ell^2 \widehat{\alpha}_\ell^2 \cos^2 (\varepsilon_\ell)}  \\
& + 2 \mathop{\sum\sum}_{1 \leq n < m \leq L} {{x_n} {x_m} {\widehat{\alpha}_n} {\widehat{\alpha}_m} {\cos(\varepsilon_n)} {\cos(\varepsilon_m)}}  ,
\end{split}
\end{equation}
\begin{equation}
\begin{split}
\Delta_{\Im}^2(\mathbf{x}) = & \; \sum_{\ell \in \mathcal{L}} {{x_\ell^2} {\widehat{\alpha}_\ell^2} {\sin^2 (\varepsilon_\ell)}} \\  
& + 2 \mathop{\sum\sum}_{1 \leq n < m \leq L} {{x_n} {x_m} {\widehat{\alpha}_n} {\widehat{\alpha}_m} {\sin(\varepsilon_n)} {\sin(\varepsilon_m)}} .
\end{split}
\end{equation}

Observe that $x_\ell^2 = x_\ell$, since $x_\ell \in \{0,1\}$. Next, by substituting the last two equations into \eqref{equation:f_d_sqrt} and then using the \emph{trigonometric identities}: $\sin^2(\vartheta) + \cos^2(\vartheta) = 1$, $\cos(\vartheta - \varphi) = \cos(\vartheta) \cos(\varphi) + \sin(\vartheta) \sin(\varphi)$, we obtain the desired result, i.e., \eqref{equation:f_d_sqrt_expanded}. Finally, the equivalence of \eqref{equation:f_d_sqrt_expanded} and \eqref{equation:f_d_sqrt_expanded_min} follows from the fact that ${x_n}{x_m} = \min({x_n},{x_m})$, for all $({x_n},{x_m}) \in \{0,1\}^2$. 
\end{IEEEproof}

\begin{remark} \label{remark:Concavity_f_d_square}
According to \eqref{equation:f_d_sqrt}, $f_\textnormal{d}^2(\mathbf{x})$ is a convex function as the sum of two convex functions: $\Delta_{\Re}^2(\mathbf{x})$ and $\Delta_{\Im}^2(\mathbf{x})$ (each one is a composition of a convex and an affine function \cite{Boyd2004}). On the other hand, if $\mu_{n m} \geq 0$ whenever $1 \leq n < m \leq L$, then, based on \eqref{equation:f_d_sqrt_expanded_min}, $f_\textnormal{d}^2(\mathbf{x})$ is a concave function since $\min({x_n},{x_m})$ is concave (the pointwise minimum of a finite set of concave functions is also a concave function \cite{Boyd2004}). In essence, by exploiting the binary nature of vector $\mathbf{x}$, we have converted the initially convex $f_\textnormal{d}^2(\mathbf{x})$ to a concave function. This is the basis of the approximation algorithm presented in Section \ref{section:Approximation_Algorithm}.  
\end{remark}

\subsection{Power Consumption Model}

The total power consumption (assuming identical IRS elements) is computed as follows
\begin{equation}
\begin{split}
\operatorname{P}_\text{tot} (\mathbf{x}) & = P_{\text{fix}} + P_{\text{on}} {L_{\text{on}} (\mathbf{x})} + P_{\text{off}} {L_{\text{off}} (\mathbf{x})}  \\
& =  P_{\text{fix}} + L P_{\text{off}} + (P_{\text{on}} - P_{\text{off}}) {\sum_{\ell \in \mathcal{L}} {x_\ell}}  ,
\end{split}
\end{equation}
where $P_{\text{fix}} = p/\eta + P_{\text{static}}$, with $\eta \in (0,1]$ being the efficiency of transmitter's power amplifier, and $P_{\text{static}} > 0$ accounting for the dissipated power in the remaining signal processing blocks at the transmitter and receiver. In addition, $L_{\text{on}} (\mathbf{x}) = \sum_{\ell \in \mathcal{L}} {x_\ell}$ and $L_{\text{off}} (\mathbf{x}) = L - L_{\text{on}} (\mathbf{x})$ are the numbers of activated and deactivated IRS elements, respectively. Furthermore, $P_{\text{on}}$, $P_{\text{off}} > 0$ represent the power consumption of each activated/deactivated element, respectively. Deactivated IRS elements usually consume a relatively small (albeit, non-negligible) amount of power compared to that of activated elements, with $P_{\text{off}} \leq P_{\text{on}}$. Similar power consumption models for passive/active IRSs can be found in \cite{Long2021} and \cite{Shojaeifard2022}. In the case of discrete phase shifts, $P_{\text{on}} = P_{\text{on}}(b)$ is monotonically increasing with $b$ \cite{Huang2019}, whereas $P_{\text{off}}$ is independent of $b$.\footnote{More accurate models for IRS power consumption are given in \cite{Wang2023} for different hardware implementations (e.g., PIN-diode-based and varactor-diode-based IRS). Those practical models can be considered for future research.}

\section{Formulation of Robust Optimization Problems Under CSI Uncertainty} \label{section:Problem_Formulation}

\subsection{Discrete IRS Phase Shifts}

The worst-case SNR is defined as the lowest SNR that can be attained under the CSI uncertainty in \eqref{equation:CSI_uncertainty}, i.e.,
\begin{subequations}  \label{problem:Worst-case_SNR_discrete}
\begin{alignat}{3}
  {\gamma}_{\text{worst}}^{\text{d}} (\mathbf{x};\delta) \! \triangleq & \mathop {\text{minimize}} \limits_{\widetilde{\mathbf{h}} \in \mathbb{C}^{L+1}} & \quad & {\gamma}^{\text{d}} (\mathbf{x},\widetilde{\mathbf{h}})  \\
& \,\text{subject to} & & {\lVert \widetilde{\mathbf{h}} \rVert}_2  \leq  \delta , \label{constraint:CSI_uncertainty}
\end{alignat}
\end{subequations}
where ${\gamma}^{\text{d}} (\mathbf{x},\widetilde{\mathbf{h}})$ is given by \eqref{equation:SNR_discrete_phase_shifts}. The global minimum of problem \eqref{problem:Worst-case_SNR_discrete} admits a \emph{closed-form expression}, according to the following theorem.\footnote{Note that a closed-form expression of the worst-case SNR is difficult to obtain in the case of multi-user multi-antenna systems, so we should construct appropriate bounds/approximations in order to formulate a robust optimization problem similar to \eqref{problem:EE_d_original}. This is a promising direction for further research.}

\begin{theorem} \label{theorem:Worst-case_SNR_discrete}
Suppose that Assumptions \ref{assumption:CSI-uncertainty_radius} and \ref{assumption:minimum_quantization_bits} hold. Then, an optimal solution to problem \eqref{problem:Worst-case_SNR_discrete} is given by
\begin{subequations}  
\begin{gather}
  \widetilde{\theta}_0^\star = (\vartheta(\mathbf{x}) + \widehat{\theta}_0 + \pi) \bmod {2\pi}  , \\
 \widetilde{\theta}_\ell^\star = (\vartheta(\mathbf{x}) + \widehat{\theta}_0 - {\phi}^{\textnormal{d}}_\ell + \pi) \bmod {2\pi} ,  \;\;\forall \ell \in \mathcal{L} , \\
  \widetilde{\alpha}_0^\star = \widetilde{\alpha}_\ell^\star = \lambda ,  \;\;\forall \ell \in \mathcal{P} , \\
 \widetilde{\alpha}_\ell^\star = 0 ,  \;\;\forall \ell \in {\mathcal{Z}} ,
\end{gather}
\end{subequations}
where $\lambda = \delta \big/ \sqrt{1 + \sum_{l \in \mathcal{L}} {x_l}}\,$, $\mathcal{P} = \{\ell \in \mathcal{L} : \, x_\ell = 1 \}$ and $\mathcal{Z} = \mathcal{L} \setminus \mathcal{P} = \{\ell \in \mathcal{L} : \, x_\ell = 0 \}$. In addition, the worst-case SNR is expressed in closed form as  
\begin{equation}  \label{equation:Worst-case_SNR_discrete} 
{\gamma}_{\textnormal{worst}}^{\textnormal{d}} (\mathbf{x};\delta) = \overline{\gamma} \left( f_{\textnormal{d}}(\mathbf{x}) - g(\mathbf{x};\delta) \right)^2,
\end{equation}
where $f_{\textnormal{d}} (\mathbf{x})$ is given by \eqref{equation:f_d}/\eqref{equation:f_d_sqrt}/\eqref{equation:f_d_sqrt_expanded}/\eqref{equation:f_d_sqrt_expanded_min}, and 
\begin{equation} \label{equation:g}
g(\mathbf{x};\delta) = \delta \sqrt{1 + \sum_{\ell \in \mathcal{L}} {x_\ell}} \, .
\end{equation}
Finally, $f_{\textnormal{d}} (\mathbf{x}) \geq g(\mathbf{x};\delta), \ \forall \mathbf{x} \in \{0,1\}^L$, and $\mu_{n m} \geq 0$ whenever $1 \leq n < m \leq L$ in \eqref{equation:f_d_sqrt_expanded}/\eqref{equation:f_d_sqrt_expanded_min}.
\end{theorem}

\begin{IEEEproof}
See Appendix \ref{appendix:Worst-case_SNR_discrete}.
\end{IEEEproof}

\noindent Based on \eqref{equation:Worst-case_SNR_discrete}, the ideal worst-case SNR (i.e., without CSI error) is equal to ${\gamma}_{\text{worst}}^\text{d,ideal} (\mathbf{x}) = {\gamma}_{\text{worst}}^\text{d} (\mathbf{x};0) = \overline{\gamma} f_\text{d}^2(\mathbf{x})$. 

Subsequently, the worst-case spectral and energy efficiency are respectively expressed as 
\begin{equation}
{\operatorname{SE}_{\text{worst}}^\text{d}}(\mathbf{x};\delta) = \log_2 \left( 1 + {\gamma_{\text{worst}}^\text{d}} (\mathbf{x};\delta) \right)  ,
\end{equation}
\begin{equation}
{\operatorname{EE}_{\text{worst}}^\text{d}}(\mathbf{x};\delta) = \frac{{\operatorname{SE}_{\text{worst}}^\text{d}}(\mathbf{x};\delta)}{\operatorname{P}_\text{tot} (\mathbf{x})}  .
\end{equation}

\noindent The \emph{robust (discrete) optimization problem} is formulated as 
\begin{subequations} \label{problem:EE_d_original}
\begin{alignat}{3}
 \operatorname{EE}_{\text{worst}}^{\text{d},\star} (\delta,\gamma_{\min}) \! \triangleq & \mathop {\text{maximize}} \limits_{\mathbf{x} \in \{ 0,1 \}^L} & \quad & {\operatorname{EE}_{\text{worst}}^\text{d}}(\mathbf{x};\delta)   \\
  & \,\text{subject to} & & {\gamma_{\text{worst}}^\text{d}} (\mathbf{x};\delta) \geq \gamma_{\min} , 
\end{alignat}
\end{subequations}
where $\gamma_{\min} \geq 0$ is the minimum required SNR. In other words, we are looking for the optimal on/off states of IRS elements in order to maximize the worst-case EE, given the CSI uncertainty, while satisfying a minimum-SNR constraint. 

A necessary and sufficient condition for feasibility as well as monotonicity properties of ${\gamma_{\text{worst}}^\text{d}} (\mathbf{x};\delta)$ and ${\operatorname{EE}_{\text{worst}}^\text{d}}(\mathbf{x};\delta)$ are provided by the following propositions. 

\begin{proposition} \label{proposition:Feasibility_discrete}
Suppose that Assumption \ref{assumption:CSI-uncertainty_radius} is true and $b \geq \nolinebreak 3$. Then, ${\gamma_{\textnormal{worst}}^\textnormal{d}} (\mathbf{x};\delta)$ is nondecreasing in each variable, i.e., ${\partial {\gamma_{\textnormal{worst}}^\textnormal{d}} (\mathbf{x};\delta)} / {\partial x_\ell} \geq 0$, $\forall \ell \in \mathcal{L}$. Moreover, problem \eqref{problem:EE_d_original} is feasible if and only if ${\gamma}_{\textnormal{worst}}^\textnormal{d} (\mathbf{1}_L;\delta) \geq \gamma_{\min}$. 
\end{proposition}

\begin{IEEEproof}
Given that Assumption \ref{assumption:CSI-uncertainty_radius} holds and because $b \geq 3$ implies Assumption \ref{assumption:minimum_quantization_bits}, Theorem \ref{theorem:Worst-case_SNR_discrete} is applicable. The partial derivative of \eqref{equation:Worst-case_SNR_discrete}, with respect to an arbitrary variable $x_\ell$, is expressed as ${\partial {\gamma_{\text{worst}}^\text{d}} (\mathbf{x};\delta)} / {\partial x_\ell} = 2 \overline{\gamma} (f_\text{d}(\mathbf{x}) - g(\mathbf{x};\delta)) \left( {\partial f_\text{d}(\mathbf{x})} / {\partial x_\ell} - {\partial g(\mathbf{x};\delta)} / {\partial x_\ell} \right)$. It suffices to show that ${\partial f_\text{d}(\mathbf{x})} / {\partial x_\ell} - {\partial g(\mathbf{x};\delta)} / {\partial x_\ell} \geq 0$, because $\overline{\gamma} \geq 0$ and $f_\text{d}(\mathbf{x}) - g(\mathbf{x};\delta) \geq 0$.

Since $b \geq 3$, we conclude that $K = 2^b \geq 8 \implies \omega = 2\pi/K \leq \pi/4 \implies |\varepsilon_\ell| \leq \omega/2 \leq \pi/8 \implies \cos(\varepsilon_\ell) \geq \cos(\pi/8) > \cos(\pi/4) = 1/\sqrt{2} \implies \zeta_\ell \geq \widehat{\alpha}_\ell^2 + \sqrt{2} \widehat{\alpha}_0 \widehat{\alpha}_\ell \geq \widehat{\alpha}_\ell^2 x_\ell + \widehat{\alpha}_0 \widehat{\alpha}_\ell$. In addition, from the triangle inequality, $ |\varepsilon_n - \varepsilon_m| \leq |\varepsilon_n| + |\varepsilon_m| \leq \omega \leq \pi/4 \implies \cos(\varepsilon_n - \varepsilon_m) \geq \cos(\pi/4) = 1/\sqrt{2} \implies \mu_{n m} \geq \sqrt{2} \widehat{\alpha}_n \widehat{\alpha}_m \geq \widehat{\alpha}_n \widehat{\alpha}_m$. Thus, the partial derivatives of \eqref{equation:f_d_sqrt_expanded} and \eqref{equation:g} are bounded by  
\begin{equation}
\begin{split}
\frac{\partial f_\text{d}(\mathbf{x})}{\partial x_\ell} & = \frac{\zeta_\ell + \sum_{n=1}^{\ell-1} {{\mu_{n\ell}} {x_n}} + \sum_{m=\ell+1}^{L} {{\mu_{\ell m}} {x_m}}}{2 f_\text{d}(\mathbf{x})}  \\ 
& \geq \frac{{\widehat{\alpha}_\ell} \left( \widehat{\alpha}_0 + \sum_{{\ell'} \in \mathcal{L}} {\widehat{\alpha}_{\ell'}} x_{\ell'} \right)}{2 f_\text{d}(\mathbf{x})} = \frac{{\widehat{\alpha}_\ell} f_\text{c}(\mathbf{x})}{2 f_\text{d}(\mathbf{x})}  \\
& \geq \frac{\widehat{\alpha}_\ell}{2} \geq \frac{\widehat{\alpha}_{\min}}{2} \geq \frac{\delta}{2}   , 
\end{split} 
\end{equation}
\begin{equation}
\frac{\partial g(\mathbf{x};\delta)}{\partial x_\ell} = \frac{\delta}{2 \sqrt{1 + \sum_{\ell \in \mathcal{L}} {x_\ell}}} \leq \frac{\delta}{2}  .
\end{equation}
In the former expression, we have taken advantage of Assumption \ref{assumption:CSI-uncertainty_radius} and the fact that $f_\text{d}(\mathbf{x}) = \left|  \widehat{\alpha}_0 + \sum_{\ell \in \mathcal{L}} {x_\ell \widehat{\alpha}_\ell e^{j\varepsilon_\ell}}  \right| \leq |\widehat{\alpha}_0| + \sum_{\ell \in \mathcal{L}} {|x_\ell \widehat{\alpha}_\ell e^{j\varepsilon_\ell}|} = \widehat{\alpha}_0 + \sum_{\ell \in \mathcal{L}} {x_\ell \widehat{\alpha}_\ell} = f_\text{c}(\mathbf{x})$, due to the triangle inequality. 

Eventually, ${\partial f_\text{d}(\mathbf{x})} / {\partial x_\ell} - {\partial g(\mathbf{x};\delta)} / {\partial x_\ell} \geq \delta/2 - \delta/2 = 0$, hence ${\partial {\gamma_{\text{worst}}^\text{d}} (\mathbf{x};\delta)} / {\partial x_\ell} \geq 0$. But $x_\ell$ is arbitrarily chosen, thus ${\partial {\gamma_{\text{worst}}^\text{d}} (\mathbf{x};\delta)} / {\partial x_\ell} \geq 0$, $\forall \ell \in \mathcal{L}$. As a result, ${\gamma}_{\textnormal{worst}}^\text{d} (\mathbf{1}_L;\delta) \geq {\gamma_{\text{worst}}^\text{d}} (\mathbf{x};\delta)$, $\forall \mathbf{x} \in \{0,1\}^L$, which easily leads to the if-and-only-if statement. 
\end{IEEEproof}

\begin{proposition} \label{proposition:Optimal_worst-case_EE_monotonicity_discrete}
Under Assumptions \ref{assumption:CSI-uncertainty_radius} and \ref{assumption:minimum_quantization_bits}, ${\gamma_{\textnormal{worst}}^\textnormal{d}} (\mathbf{x};\delta)$ and ${\operatorname{EE}_{\textnormal{worst}}^\textnormal{d}}(\mathbf{x};\delta)$ are nonincreasing functions of $\delta$, i.e., ${\partial {\gamma_{\textnormal{worst}}^\textnormal{d}} (\mathbf{x};\delta)} / {\partial \delta} \leq 0$ and ${\partial {\operatorname{EE}_{\textnormal{worst}}^\textnormal{d}}(\mathbf{x};\delta)} / {\partial \delta} \leq 0$. Furthermore, the optimal worst-case EE, defined by \eqref{problem:EE_d_original}, is nonincreasing in each of its arguments, i.e., 
\begin{equation}  
\begin{split}
  \delta_1 \leq \delta_2 \implies & \operatorname{EE}_{\textnormal{worst}}^{\textnormal{d},\star} (\delta_1,\gamma_{\min}) \geq \operatorname{EE}_{\textnormal{worst}}^{\textnormal{d},\star} (\delta_2,\gamma_{\min}) , \\
  & \; \forall \gamma_{\min} \geq 0  , 
\end{split}  
\end{equation}
\begin{equation}
\begin{split}
  \gamma_{\min}^{(1)} \leq \gamma_{\min}^{(2)} \implies & \operatorname{EE}_{\textnormal{worst}}^{\textnormal{d},\star} (\delta,\gamma_{\min}^{(1)}) \geq \operatorname{EE}_{\textnormal{worst}}^{\textnormal{d},\star} (\delta,\gamma_{\min}^{(2)}) ,  \\
  & \; \forall \delta \in \mathcal{A}  ,
\end{split}
\end{equation}
where $\mathcal{A} = \{ \delta \in \mathbb{R} : \, 0 \leq \delta \leq \widehat{\alpha}_{\min} \}$.
\end{proposition}

\begin{IEEEproof}
First of all, Theorem \ref{theorem:Worst-case_SNR_discrete} is applicable because of Assumptions \ref{assumption:CSI-uncertainty_radius} and \ref{assumption:minimum_quantization_bits}. Now, let $\mathcal{F}(\delta,\gamma_{\min}) = \{\mathbf{x} \in \{ 0,1 \}^L : \, {\gamma_{\text{worst}}^\text{d}} (\mathbf{x};\delta) \geq \gamma_{\min} \}$ denote the feasible set of problem \eqref{problem:EE_d_original}. Moreover, we have ${\partial g(\mathbf{x};\delta)} / {\partial \delta} = \sqrt{1 + \sum_{\ell \in \mathcal{L}} {x_\ell}} \geq 1 > 0 \implies {\partial {\gamma_{\text{worst}}^\text{d}} (\mathbf{x};\delta)} / {\partial \delta} = - 2\overline{\gamma} (f_\text{d}(\mathbf{x}) - g(\mathbf{x};\delta)) {{\partial g(\mathbf{x};\delta)} / {\partial \delta}} \leq 0 \implies {\partial {\operatorname{EE}_{\text{worst}}^\text{d}}(\mathbf{x};\delta)} / {\partial \delta} = [\log(2) {\operatorname{P}_\text{tot} (\mathbf{x})} (1 + {\gamma_{\text{worst}}^\text{d}} (\mathbf{x};\delta))]^{-1} {\partial {\gamma_{\text{worst}}^\text{d}} (\mathbf{x};\delta)} / {\partial \delta} \leq 0$, i.e., ${\gamma_{\text{worst}}^\text{d}} (\mathbf{x};\delta)$ and ${\operatorname{EE}_{\text{worst}}^\text{d}}(\mathbf{x};\delta)$ are nonincreasing functions of $\delta$. Consequently, we obtain the following implications: 1) $\delta_1 \leq \delta_2 \implies [{\gamma_{\text{worst}}^\text{d}} (\mathbf{x};\delta_1) \geq {\gamma_{\text{worst}}^\text{d}} (\mathbf{x};\delta_2)] \land [{\operatorname{EE}_{\text{worst}}^\text{d}}(\mathbf{x};\delta_1) \geq {\operatorname{EE}_{\text{worst}}^\text{d}}(\mathbf{x};\delta_2)] \land [\mathcal{F}(\delta_2,\gamma_{\min}) \subseteq \mathcal{F}(\delta_1,\gamma_{\min})] \implies \operatorname{EE}_{\textnormal{worst}}^{\text{d},\star} (\delta_1,\gamma_{\min}) = \max_{\mathbf{x} \in \mathcal{F}(\delta_1,\gamma_{\min})} {{\operatorname{EE}_{\text{worst}}^\text{d}}(\mathbf{x};\delta_1)} \geq \max_{\mathbf{x} \in \mathcal{F}(\delta_2,\gamma_{\min})} {{\operatorname{EE}_{\text{worst}}^\text{d}}(\mathbf{x};\delta_1)} \geq \max_{\mathbf{x} \in \mathcal{F}(\delta_2,\gamma_{\min})} {{\operatorname{EE}_{\text{worst}}^\text{d}}(\mathbf{x};\delta_2)} = \operatorname{EE}_{\textnormal{worst}}^{\text{d},\star} (\delta_2,\gamma_{\min})$, and 2) $\gamma_{\min}^{(1)} \leq \gamma_{\min}^{(2)} \implies \mathcal{F}(\delta,\gamma_{\min}^{(2)}) \subseteq \mathcal{F}(\delta,\gamma_{\min}^{(1)}) \implies \operatorname{EE}_{\textnormal{worst}}^{\text{d},\star} (\delta,\gamma_{\min}^{(1)}) = \max_{\mathbf{x} \in \mathcal{F}(\delta,\gamma_{\min}^{(1)})} {{\operatorname{EE}_{\text{worst}}^\text{d}}(\mathbf{x};\delta)} \geq \max_{\mathbf{x} \in \mathcal{F}(\delta,\gamma_{\min}^{(2)})} {{\operatorname{EE}_{\text{worst}}^\text{d}}(\mathbf{x};\delta)} = \operatorname{EE}_{\textnormal{worst}}^{\text{d},\star} (\delta,\gamma_{\min}^{(2)})$. 
\end{IEEEproof}

\subsection{Continuous IRS Phase Shifts}

In a similar way, the worst-case SNR for continuous phase shifts is defined as follows 
\begin{subequations}  \label{problem:Worst-case_SNR_continuous}
\begin{alignat}{3}
{\gamma}_{\text{worst}}^\text{c} (\mathbf{x};\delta) \! \triangleq & \mathop {\text{minimize}} \limits_{\widetilde{\mathbf{h}} \in \mathbb{C}^{L+1}} & \quad & {\gamma^\text{c}} (\mathbf{x},\widetilde{\mathbf{h}})  \\
& \,\text{subject to} & & {\lVert \widetilde{\mathbf{h}} \rVert}_2  \leq  \delta . 
\end{alignat}
\end{subequations} 
where ${\gamma^\text{c}} (\mathbf{x},\widetilde{\mathbf{h}})$ is given by \eqref{equation:SNR_continuous_phase_shifts}. The \emph{robust EE-maximization problem} is now formulated as 
\begin{subequations} \label{problem:EE_c_original}
\begin{alignat}{3}
 \operatorname{EE}_{\text{worst}}^{\text{c},\star} (\delta,\gamma_{\min}) \! \triangleq & \mathop {\text{maximize}} \limits_{\mathbf{x} \in \{ 0,1 \}^L} & \quad & {\operatorname{EE}_{\text{worst}}^\text{c}}(\mathbf{x};\delta)   \\
  & \,\text{subject to} & & {\gamma_{\text{worst}}^\text{c}} (\mathbf{x};\delta) \geq \gamma_{\min} , 
\end{alignat}
\end{subequations}
where $\gamma_{\min} \geq 0$ is the minimum required SNR, and  
\begin{equation}
{\operatorname{SE}_{\text{worst}}^\text{c}}(\mathbf{x};\delta) = \log_2 \left( 1 + {\gamma_{\text{worst}}^\text{c}} (\mathbf{x};\delta) \right)  ,
\end{equation}
\begin{equation}
{\operatorname{EE}_{\text{worst}}^\text{c}}(\mathbf{x};\delta) = \frac{{\operatorname{SE}_{\text{worst}}^\text{c}}(\mathbf{x};\delta)}{\operatorname{P}_\text{tot} (\mathbf{x})}  .
\end{equation}

For the sake of completeness, we present the ``continuous'' counterparts of Theorem \ref{theorem:Worst-case_SNR_discrete} and Propositions \ref{proposition:Feasibility_discrete}, \ref{proposition:Optimal_worst-case_EE_monotonicity_discrete}. 

\begin{remark}
Due to Proposition \ref{proposition:limits}, continuous phase shifts can be seen as a special case of discrete phase shifts by making the following substitutions: $\vartheta \mapsto 0$, ${\phi}^{\textnormal{d}}_\ell \mapsto \phi_\ell^\star$, for all $\ell \in \mathcal{L}$, $\gamma^\textnormal{d} \mapsto \gamma^\textnormal{c}$, $f_\textnormal{d} \mapsto f_\textnormal{c}$, $\operatorname{EE}^\textnormal{d} \mapsto \operatorname{EE}^\textnormal{c}$. Finally, Assumption \ref{assumption:minimum_quantization_bits} and $b \geq 3$ can be removed, since both conditions are automatically satisfied when $b$ tends to infinity. Detailed proofs can be found in the conference paper \cite{Efrem2023_ICC}. 
\end{remark}

\begin{theorem} \label{theorem:Worst-case_SNR_continuous}
Suppose that Assumption \ref{assumption:CSI-uncertainty_radius} holds. Then, an optimal solution to problem \eqref{problem:Worst-case_SNR_continuous} is given by
\begin{subequations}  
\begin{gather}
  \widetilde{\theta}_0^\star = (\widehat{\theta}_0 + \pi) \bmod {2\pi}  , \\
 \widetilde{\theta}_\ell^\star = (\widehat{\theta}_0 - {\phi}_\ell^{\star} + \pi) \bmod {2\pi} ,  \;\;\forall \ell \in \mathcal{L} , \\
  \widetilde{\alpha}_0^\star = \widetilde{\alpha}_\ell^\star = \lambda ,  \;\;\forall \ell \in \mathcal{P} , \\
 \widetilde{\alpha}_\ell^\star = 0 ,  \;\;\forall \ell \in {\mathcal{Z}} ,
\end{gather}
\end{subequations}
where $\lambda = \delta \big/ \sqrt{1 + \sum_{l \in \mathcal{L}} {x_l}}\,$, $\mathcal{P} = \{\ell \in \mathcal{L} : \, x_\ell = 1 \}$ and $\mathcal{Z} = \mathcal{L} \setminus \mathcal{P} = \{\ell \in \mathcal{L} : \, x_\ell = 0 \}$. In addition, the worst-case SNR is expressed in closed form as  
\begin{equation}  \label{equation:Worst-case_SNR_continuous} 
{\gamma}_{\textnormal{worst}}^\textnormal{c} (\mathbf{x};\delta) = \overline{\gamma} \left( f_\textnormal{c}(\mathbf{x}) - g(\mathbf{x};\delta) \right)^2,
\end{equation}
where $f_\textnormal{c}(\mathbf{x})$ and $g(\mathbf{x};\delta)$ are given by \eqref{equation:f_c} and \eqref{equation:g}, respectively. Finally, $f_\textnormal{c}(\mathbf{x}) \geq g(\mathbf{x};\delta), \ \forall \mathbf{x} \in \{0,1\}^L$.
\end{theorem}

\noindent Based on \eqref{equation:Worst-case_SNR_continuous}, the ideal worst-case SNR (i.e., without CSI error) is equal to ${\gamma}_{\text{worst}}^\text{c,ideal} (\mathbf{x}) = {\gamma}_{\text{worst}}^\text{c} (\mathbf{x};0) = \overline{\gamma} f_\text{c}^2(\mathbf{x})$.

\begin{proposition} \label{proposition:Feasibility_continuous}
Suppose that Assumption \ref{assumption:CSI-uncertainty_radius} is true. Then, ${\gamma_{\textnormal{worst}}^\textnormal{c}} (\mathbf{x};\delta)$ is nondecreasing in each variable, i.e., ${\partial {\gamma_{\textnormal{worst}}^\textnormal{c}} (\mathbf{x};\delta)} / {\partial x_\ell} \geq 0$, $\forall \ell \in \mathcal{L}$. Moreover, problem \eqref{problem:EE_c_original} is feasible if and only if ${\gamma}_{\textnormal{worst}}^\textnormal{c} (\mathbf{1}_L;\delta) \geq \gamma_{\min}$.
\end{proposition}

\begin{proposition} \label{proposition:Optimal_worst-case_EE_monotonicity_continuous}
Under Assumption \ref{assumption:CSI-uncertainty_radius}, ${\gamma_{\textnormal{worst}}^\textnormal{c}} (\mathbf{x};\delta)$ and ${\operatorname{EE}_{\textnormal{worst}}^\textnormal{c}}(\mathbf{x};\delta)$ are nonincreasing functions of $\delta$, i.e., ${\partial {\gamma_{\textnormal{worst}}^\textnormal{c}} (\mathbf{x};\delta)} / {\partial \delta} \leq 0$ and ${\partial {\operatorname{EE}_{\textnormal{worst}}^\textnormal{c}}(\mathbf{x};\delta)} / {\partial \delta} \leq 0$. Furthermore, the optimal worst-case EE, defined by \eqref{problem:EE_c_original}, is nonincreasing in each of its arguments, i.e., 
\begin{equation}  
\begin{split}
  \delta_1 \leq \delta_2 \implies & \operatorname{EE}_{\textnormal{worst}}^{\textnormal{c},\star} (\delta_1,\gamma_{\min}) \geq \operatorname{EE}_{\textnormal{worst}}^{\textnormal{c},\star} (\delta_2,\gamma_{\min}) , \\
  & \; \forall \gamma_{\min} \geq 0  , 
\end{split}  
\end{equation}
\begin{equation}
\begin{split}
  \gamma_{\min}^{(1)} \leq \gamma_{\min}^{(2)} \implies & \operatorname{EE}_{\textnormal{worst}}^{\textnormal{c},\star} (\delta,\gamma_{\min}^{(1)}) \geq \operatorname{EE}_{\textnormal{worst}}^{\textnormal{c},\star} (\delta,\gamma_{\min}^{(2)}) ,  \\
  & \; \forall \delta \in \mathcal{A}  ,
\end{split}
\end{equation}
where $\mathcal{A} = \{ \delta \in \mathbb{R} : \, 0 \leq \delta \leq \widehat{\alpha}_{\min} \}$.
\end{proposition}

\begin{remark}
The robust optimization problems \eqref{problem:EE_d_original} and \eqref{problem:EE_c_original} can be globally solved using \emph{exhaustive search}, however, with \emph{exponential complexity} $\Theta \left( \sum_{k=0}^L {\binom{L}{k} k} \right) = \Theta ({2^L}L)$. Therefore, exhaustive search is impractical for relatively large $L$, which is usually the case.
\end{remark}

\section{Global Optimization Using Dynamic Programming -- Continuous Phase Shifts}  \label{section:Dynamic_Programming}

In this section, we design a \emph{dynamic programming (DP) algorithm} that achieves the \emph{global optimum of problem \eqref{problem:EE_c_original} in polynomial time} $O(L\,{\log L})$.\footnote{A DP algorithm was also designed in \cite{Efrem2021} for the selection of ground stations in satellite systems so as to minimize the total installation cost.} Thus, the problem is computationally tractable despite being nonconvex.

Subsequently, problem \eqref{problem:EE_c_original} can be decomposed into $L+1$ subproblems of the following form, where $M \in \mathcal{L}_0$, 
\begin{subequations}  \label{subproblem:equal_M}
\begin{alignat}{3}
  \operatorname{EE}^\star_{=M} \! \triangleq  & \mathop {\text{maximize}} \limits_{\mathbf{x} \in \{ 0,1 \}^L} & \quad & {\operatorname{EE}_\text{worst}^\text{c}}(\mathbf{x};\delta)  \\
  & \,\text{subject to} & & {\gamma_\text{worst}^\text{c}}(\mathbf{x};\delta) \geq \gamma_{\min} , \\
  & & & \sum_{\ell \in \mathcal{L}} {x_\ell} = M  .
\end{alignat}
\end{subequations}
Henceforth, we will write $\operatorname{EE}_{\text{worst}}^{\text{c},\star}$ instead of $\operatorname{EE}_{\text{worst}}^{\text{c},\star} (\delta,\gamma_{\min})$, and make the convention that if a maximization problem is not feasible, then its optimal value is set to $-\infty$.

\begin{remark} \label{remark:Subproblems_decomposition}
The decomposition of \eqref{problem:EE_c_original} is essentially a partition of its feasible set, thus $\operatorname{EE}^{\textnormal{c},\star}_\textnormal{worst} = \max_{M \in \mathcal{L}_0} \{ \operatorname{EE}^\star_{=M} \}$. In addition, problem \eqref{problem:EE_c_original} is feasible if and only if there exists an $M \in \mathcal{L}_0$ such that subproblem \eqref{subproblem:equal_M} is feasible. 
\end{remark}

According to the following proposition, problem \eqref{subproblem:equal_M} can be solved by selecting $M$ IRS elements with the largest $\widehat{\alpha}_\ell$.

\begin{proposition} \label{proposition:Equivalent_subproblem}
Suppose that Assumption \ref{assumption:CSI-uncertainty_radius} is true. Then, for any given $M \in \mathcal{L}_0$, subproblem \eqref{subproblem:equal_M} is equivalent to\footnote{We say that two optimization problems are ``equivalent'' if they have equal sets of optimal solutions, i.e., any optimal solution to the first problem is also an optimal solution to the second problem and vice versa.}
\begin{subequations}  \label{subproblem:equal_M_equivalent}
\begin{alignat}{3}
 & \mathop {\textnormal{maximize}} \limits_{\mathbf{x} \in \{ 0,1 \}^L} & \quad & f_{\textnormal{c}}(\mathbf{x}) \\
  & \,\textnormal{subject to} & & f_{\textnormal{c}}(\mathbf{x}) \geq f_{\min}(M) , \\
  & & & \sum_{\ell \in \mathcal{L}} {x_\ell} = M  ,
\end{alignat}
\end{subequations}
where $f_{\min}(M) = \sqrt{{\gamma_{\min}}/{\overline{\gamma}}} + \delta \sqrt{1+M}$. Moreover, subproblem \eqref{subproblem:equal_M}/\eqref{subproblem:equal_M_equivalent} is feasible if and only if $\widehat{\alpha}_0 + \sum_{m=1}^{M} {\widehat{\alpha}_{\zeta_m}} \geq f_{\min}(M)$, where $(\zeta_1,\dots,\zeta_L)$ is a permutation of $\mathcal{L}$ such that $\widehat{\alpha}_{\zeta_1} \geq \cdots \geq \widehat{\alpha}_{\zeta_L}$. In addition, an optimal solution to subproblem \eqref{subproblem:equal_M}/\eqref{subproblem:equal_M_equivalent}, provided that it is feasible, is given by: $x_{\zeta_m}^\star = 1$, $\forall m \in \mathcal{M}=\{1,\dots,M\}$, and $x_{\zeta_\ell}^\star = 0$, $\forall \ell \in {\mathcal{L} \setminus \mathcal{M}}$. 
\end{proposition}

\begin{IEEEproof}
Based on Theorem \ref{theorem:Worst-case_SNR_continuous}, a key observation is that, for fixed $\sum_{\ell \in \mathcal{L}} {x_\ell} = M$, the worst-case SNR and total power consumption are written as $\gamma^\text{c}_\text{worst} (\mathbf{x};\delta) = \overline{\gamma} \left( f_{\text{c}}(\mathbf{x}) - \delta \sqrt{1+M} \right)^2$ and $\operatorname{P}_\text{tot} (\mathbf{x}) =  P_{\text{fix}} + L P_{\text{off}} + (P_{\text{on}} - P_{\text{off}}) M$. As a result, we obtain the problem equivalence, since $\log_2(y)$ and $y^2$ are both increasing functions for $y>0$. The if-and-only-if statement about feasibility and the optimal solution follow directly from the fact that $\max_{\mathbf{x} \in \{ 0,1 \}^L} \left\{ f_{\text{c}}(\mathbf{x}) : \, \sum_{\ell \in \mathcal{L}} {x_\ell} = M \right\} = \widehat{\alpha}_0 + \sum_{m=1}^{M} {\widehat{\alpha}_{\zeta_m}}$. 
\end{IEEEproof}

\noindent Also, we define the following subproblem, for any $M \in \mathcal{L}_0$, 
\begin{subequations}  \label{subproblem:less_equal_M}
\begin{alignat}{3}
  \operatorname{EE}^\star_{\leq M} \! \triangleq  & \mathop {\text{maximize}} \limits_{\mathbf{x} \in \{ 0,1 \}^L} & \quad & {\operatorname{EE}_\text{worst}^\text{c}}(\mathbf{x};\delta)  \\
  & \,\text{subject to} & & {\gamma_\text{worst}^\text{c}}(\mathbf{x};\delta) \geq \gamma_{\min} , \\
  & & & \sum_{\ell \in \mathcal{L}} {x_\ell} \leq M  .
\end{alignat} 
\end{subequations}
This subproblem will facilitate the identification of the \linebreak \emph{optimal substructure}, which is a fundamental ingredient of DP \cite{Cormen2009}.\footnote{An optimization problem is said to have ``optimal substructure'' if its optimal value can be computed using the optimal values of its subproblems.} Since the constraint $\sum_{\ell \in \mathcal{L}} {x_\ell} \leq M$ is equivalent to $\left( \sum_{\ell \in \mathcal{L}} {x_\ell} \leq M-1 \right) \lor \left( \sum_{\ell \in \mathcal{L}} {x_\ell} = M \right)$, we have the following \emph{recurrence relation}  
\begin{equation} \label{equation:Recurrence_relation}
\operatorname{EE}_{\leq {M}}^\star = \max({\operatorname{EE}_{\leq M-1}^\star},{\operatorname{EE}_{=M}^\star}), \;\; \forall M \in \mathcal{L} ,
\end{equation}
with initial condition $\operatorname{EE}_{\leq 0}^\star = \operatorname{EE}_{=0}^\star$. 

\begin{remark} \label{remark:Recurrence_relation}
$\left\{ \operatorname{EE}_{=M}^\star \right\}_{M \in \mathcal{L}_0}$ can be computed by exploiting Proposition \ref{proposition:Equivalent_subproblem}. Moreover, $\operatorname{EE}_{\leq L}^\star = \operatorname{EE}^{\textnormal{c},\star}_\textnormal{worst}$ because the constraint $\sum_{\ell \in \mathcal{L}} {x_\ell} \leq L$ holds for all $\mathbf{x} \in \{0,1\}^L$, so it can be omitted. In other words, subproblem \eqref{subproblem:less_equal_M} for $M=L$ is identical to problem \eqref{problem:EE_c_original}. We can also prove, using mathematical induction, that the solution of \eqref{equation:Recurrence_relation} is $\operatorname{EE}^{\textnormal{c},\star}_\textnormal{worst} = \operatorname{EE}_{\leq L}^\star = \max_{M \in \mathcal{L}_0} \{ \operatorname{EE}^\star_{=M} \}$, which is in agreement with Remark \ref{remark:Subproblems_decomposition}. 
\end{remark}

The DP approach is presented in Algorithm \ref{algorithm:DP}. First, the algorithm sorts the entries of $[{\widehat{\alpha}_1}, \ldots ,{\widehat{\alpha}_L}]^\top$ in descending order and initializes some programming variables (steps 1--8). Subsequently, the for-loop (steps 9--15) updates the values of $\operatorname{EE}^\star$ and $M^\star$ according to the recurrence relation \eqref{equation:Recurrence_relation}. Finally, steps 16--24 decide the feasibility of problem \eqref{problem:EE_c_original}, and reconstruct an optimal solution if the problem is feasible. More precisely, the correctness and polynomial complexity of the proposed algorithm are established by the following theorem.

\begin{theorem} \label{theorem:DP_algorithm}
Provided that Assumption \ref{assumption:CSI-uncertainty_radius} holds, Algorithm \ref{algorithm:DP} returns either \textit{``Infeasible''} if problem \eqref{problem:EE_c_original} is not feasible, or its global maximum together with an optimal solution otherwise. Furthermore, the complexity of Algorithm \ref{algorithm:DP} is $O(L\,{\log L})$. 
\end{theorem}

\begin{IEEEproof}
See Appendix \ref{appendix:DP_algorithm}.
\end{IEEEproof}

\noindent
\begin{minipage}[!t]{\columnwidth}
\begin{algorithm}[H]   
\caption{Dynamic Programming (DP)}  \label{algorithm:DP}
\small
\begin{algorithmic}[1] 
\State Sort the entries of $[{\widehat{\alpha}_1}, \ldots ,{\widehat{\alpha}_L}]^\top$ in descending order. Let $({\zeta_1}, \ldots ,{\zeta_L})$ be a permutation of $\mathcal{L}$ such that $\widehat{\alpha}_{\zeta_1} \geq \cdots \geq \widehat{\alpha}_{\zeta_L}$.
\State $f_\text{c} :=  \widehat{\alpha}_0$, $\gamma^\text{c}_\text{worst} := \overline{\gamma} \left( f_\text{c} - \delta \right)^2$ 
\State $\operatorname{P}_\text{tot} :=  P_{\text{fix}} + L P_{\text{off}}$, $\operatorname{\Delta P} := P_{\text{on}} - P_{\text{off}}$ 
\If{$\gamma^\text{c}_\text{worst} \geq \gamma_{\min}$}
	\State  ${\operatorname{EE}^\star} := \log_2(1+\gamma^\text{c}_\text{worst})/\operatorname{P}_\text{tot}$, $M^\star := 0$ 
\Else 
	\State ${\operatorname{EE}^\star} := -\infty$ 
\EndIf 
\For{$M : = 1\ \text{to}\ L$}
	\State $f_\text{c} := f_\text{c} + \widehat{\alpha}_{\zeta_M}$, $\gamma^\text{c}_\text{worst} := \overline{\gamma} \left( f_\text{c} - \delta \sqrt{1+M} \right)^2$
	\State $\operatorname{P}_\text{tot} :=  \operatorname{P}_\text{tot} + \operatorname{\Delta P}$, $\operatorname{EE} := \log_2(1+\gamma^\text{c}_\text{worst})/\operatorname{P}_\text{tot}$ 
	\If{$(\gamma^\text{c}_\text{worst} \geq \gamma_{\min}) \land (\operatorname{EE} > \operatorname{EE}^\star)$}
		\State $\operatorname{EE}^\star := \operatorname{EE}$, $M^\star := M$    
	\EndIf	
\EndFor 
\If{${\operatorname{EE}^\star} = -\infty$} 
	\State \textbf{return} \textit{``Infeasible''} 
\Else
	\State $\mathbf{x}^\star := \mathbf{0}_L$
	\For{$m := 1\ \text{to}\ M^\star$}
		\State $x_{\zeta_m}^\star := 1$
	\EndFor
	\State \textbf{return} $({\operatorname{EE}^\star},{\mathbf{x}^\star})$		
\EndIf
\end{algorithmic}
\end{algorithm} 
\end{minipage}

\section{Convex-Relaxation-Based Method -- \\ Discrete Phase Shifts} \label{section:Approximation_Algorithm}

In this section, we design and analyze a polynomial-time approximation algorithm, based on convex relaxation, in order to tackle problem \eqref{problem:EE_d_original}. This problem is quite different from problem \eqref{problem:EE_c_original}, mainly because $f_\text{d} (\mathbf{x})$ is \emph{not} an affine function like $f_\text{c} (\mathbf{x})$. As a result, global optimization in the case of discrete phase shifts is no longer computationally tractable compared to the case of continuous phase shifts. 

First of all, by virtue of Assumptions \ref{assumption:CSI-uncertainty_radius} and \ref{assumption:minimum_quantization_bits}, the worst-case SNR in \eqref{equation:Worst-case_SNR_discrete} can be upper bounded as follows\footnote{Under Assumptions \ref{assumption:CSI-uncertainty_radius} and \ref{assumption:minimum_quantization_bits}, Theorem \ref{theorem:Worst-case_SNR_discrete} implies that $f_{\textnormal{d}} (\mathbf{x}) \geq g(\mathbf{x};\delta)$.} 
\begin{equation}
\begin{split}
{\gamma}_{\textnormal{worst}}^{\textnormal{d}} (\mathbf{x};\delta) & = \overline{\gamma} \left( f_{\textnormal{d}}(\mathbf{x}) - g(\mathbf{x};\delta) \right)^2  \\
& \leq \overline{\gamma} \left( f_{\textnormal{d}}(\mathbf{x}) - g(\mathbf{x};\delta) \right) \left( f_{\textnormal{d}}(\mathbf{x}) + g(\mathbf{x};\delta) \right)     \\
& =  \overline{\gamma} \left( f_{\textnormal{d}}^2(\mathbf{x}) - g^2(\mathbf{x};\delta) \right)   \\
& =  \overline{\gamma} \left( \xi + \sum_{\ell \in \mathcal{L}}{{\zeta'_\ell} {x_\ell}} + \mathop{\sum\sum}_{1 \leq n < m \leq L} {{\mu_{n m}} {\min({x_n},{x_m})}} \right)   \\
&  \triangleq \widehat{\gamma}_{\textnormal{worst}}^{\textnormal{d}} (\mathbf{x};\delta)  ,
\end{split} 
\end{equation} 
where $\xi = \widehat{\alpha}_0^2 - \delta^2$, and $\zeta'_\ell = \zeta_\ell - \delta^2$ for all $\ell \in \mathcal{L}$. In a similar manner, we can upper bound the worst-case EE, i.e., 
\begin{equation}
{\operatorname{EE}_{\text{worst}}^\text{d}}(\mathbf{x};\delta) = \frac{{\operatorname{SE}_{\text{worst}}^\text{d}}(\mathbf{x};\delta)}{\operatorname{P}_\text{tot} (\mathbf{x})}  \leq  \frac{\widehat{\operatorname{SE}}\vphantom{\widehat{\operatorname{SE}}}_{\text{worst}}^\text{d} (\mathbf{x};\delta)}{\operatorname{P}_\text{tot} (\mathbf{x})} \triangleq  \widehat{\operatorname{EE}}\vphantom{\widehat{\operatorname{EE}}}_{\text{worst}}^\text{d} (\mathbf{x};\delta)  ,  
\end{equation}
where $\widehat{\operatorname{SE}}\vphantom{\widehat{\operatorname{SE}}}_{\text{worst}}^\text{d} (\mathbf{x};\delta) = \log_2 \left( 1 + \widehat{\gamma}_{\text{worst}}^\text{d} (\mathbf{x};\delta) \right)$. Observe that the above bounds are tight as $\delta$ tends to 0, i.e., $\lim_{\delta \to 0} {\gamma}_{\textnormal{worst}}^{\textnormal{d}} (\mathbf{x};\delta) = \lim_{\delta \to 0} \widehat{\gamma}_{\textnormal{worst}}^{\textnormal{d}} (\mathbf{x};\delta) = \overline{\gamma} f_{\textnormal{d}}^2(\mathbf{x})$, and   $\lim_{\delta \to 0} {\operatorname{EE}_{\text{worst}}^\text{d}}(\mathbf{x};\delta) = \lim_{\delta \to 0} \widehat{\operatorname{EE}}\vphantom{\widehat{\operatorname{EE}}}_{\text{worst}}^\text{d} (\mathbf{x};\delta) = {\log_2 \left( 1 + \overline{\gamma} f_{\textnormal{d}}^2(\mathbf{x}) \right)} / {\operatorname{P}_\text{tot} (\mathbf{x})}$. 

According to Theorem \ref{theorem:Worst-case_SNR_discrete} and Remark \ref{remark:Concavity_f_d_square}, $f_\textnormal{d}^2(\mathbf{x})$ is concave in $\mathbf{x}$, and so is $\widehat{\gamma}_{\textnormal{worst}}^{\textnormal{d}} (\mathbf{x};\delta)$. In addition, $\widehat{\operatorname{SE}}\vphantom{\widehat{\operatorname{SE}}}_{\text{worst}}^\text{d} (\mathbf{x};\delta)$ is concave in $\mathbf{x}$, since $\log_2(1+y)$ is concave and nondecreasing \cite{Boyd2004}. Therefore, $\widehat{\operatorname{EE}}\vphantom{\widehat{\operatorname{EE}}}_{\text{worst}}^\text{d} (\mathbf{x};\delta)$ is a concave-affine (thus, concave-convex) ratio of $\mathbf{x}$. Consequently, we formulate a \emph{concave-affine fractional relaxation problem} (by also relaxing the binary constraints)
\begin{subequations} \label{problem:Fractional_relaxation}
\begin{alignat}{3}
 \operatorname{EE}_{\text{rel}}^{\star} (\delta,\gamma_{\min}) \! \triangleq & \mathop {\text{maximize}} \limits_{\mathbf{x} \in [ 0,1 ]^L} & \quad & \widehat{\operatorname{EE}}\vphantom{\widehat{\operatorname{EE}}}_{\text{worst}}^\text{d} (\mathbf{x};\delta)   \\
  & \,\text{subject to} & & \widehat{\gamma}_{\textnormal{worst}}^{\textnormal{d}} (\mathbf{x};\delta) \geq \gamma_{\min} .
\end{alignat}
\end{subequations}

\begin{remark} \label{remark:Fractional_relaxation}
Let $\mathcal{F}$ and $\mathcal{F}_\textnormal{rel}$ be the feasible sets of problems \eqref{problem:EE_d_original} and \eqref{problem:Fractional_relaxation}, respectively. It can be easily proved that $\mathcal{F} \subseteq \mathcal{F}_\textnormal{rel}$, and therefore $\operatorname{EE}_{\textnormal{worst}}^{\textnormal{d},\star} (\delta,\gamma_{\min}) = \max_{\mathbf{x} \in \mathcal{F}} {{\operatorname{EE}_{\textnormal{worst}}^\textnormal{d}}(\mathbf{x};\delta)} \leq \max_{\mathbf{x} \in \mathcal{F}_\textnormal{rel}} {{\operatorname{EE}_{\textnormal{worst}}^\textnormal{d}}(\mathbf{x};\delta)} \leq   \max_{\mathbf{x} \in \mathcal{F}_\textnormal{rel}} {\widehat{\operatorname{EE}}\vphantom{\widehat{\operatorname{EE}}}_{\textnormal{worst}}^\textnormal{d} (\mathbf{x};\delta)} = \operatorname{EE}_{\textnormal{rel}}^{\star} (\delta,\gamma_{\min})$. In addition, if problem \eqref{problem:EE_d_original} is feasible (i.e., $\mathcal{F} \neq \varnothing$), then so is the relaxation problem \eqref{problem:Fractional_relaxation}, i.e., $\mathcal{F}_\textnormal{rel} \neq \varnothing$. 
\end{remark}

Problem \eqref{problem:Fractional_relaxation} can be globally solved using either \emph{Dinkelbach's iterative algorithm} \cite{Dinkelbach1967} or \emph{Charnes-Cooper's transform} \cite{Charnes1962,Schaible1974}. The former method achieves the global optimum by solving a sequence of convex problems, without introducing extra variables and constraints, while the latter method solves a single equivalent convex problem with additional variable and constraint \cite{Zappone2015}. 

In this paper, we adopt the Charnes-Cooper transform. First, we introduce an additional variable $t = 1/{\operatorname{P}_\text{tot} (\mathbf{x})} > 0$, and then apply a (component-wise) change of variables, i.e., $\mathbf{y} = \mathbf{x}/{\operatorname{P}_\text{tot} (\mathbf{x})} = t \, \mathbf{x}$, where $\mathbf{y}=[{y_1},\ldots,{y_L}]^\top$. In this way, problem \eqref{problem:Fractional_relaxation} is equivalent to the following \emph{convex problem}
\begin{subequations} \label{problem:Convex_relaxation}
\begin{alignat}{3}
\operatorname{EE}_{\text{rel}}^{\star} (\delta,\gamma_{\min}) \! = & \mathop {\text{maximize}} \limits_{\mathbf{y},t} & \quad & t \, \widehat{\operatorname{SE}}\vphantom{\widehat{\operatorname{SE}}}_{\text{worst}}^\text{d} (\mathbf{y}/t;\delta)   \\
  & \,\text{subject to} & & t \, {\widehat{\gamma}_{\textnormal{worst}}^{\textnormal{d}} (\mathbf{y}/t;\delta)} \geq t \, {\gamma_{\min}}  ,  \\ 
  & & & {0 \leq y_\ell \leq t},\;\;\forall \ell \in \mathcal{L} ,  \\
  & & & t \, {\operatorname{P}_\text{tot} (\mathbf{y}/t)} = 1  .  
\end{alignat}
\end{subequations}
Recall that the perspective operation preserves convexity and concavity: if $f(\mathbf{x})$ is convex (or concave), then so is its perspective function $g(\mathbf{y},t) = t {f(\mathbf{y}/t)}$ for $t > 0$ \cite{Boyd2004}. In particular,  
\begin{equation}
t \, {\operatorname{P}_\text{tot} (\mathbf{y}/t)} = (P_{\text{fix}} + L P_{\text{off}})t + (P_{\text{on}} - P_{\text{off}}) \sum_{\ell \in \mathcal{L}} {y_\ell} 
\end{equation}
is linear (hence, affine), and
\begin{equation}
\scalebox{0.95}{$
t \, {\widehat{\gamma}_{\textnormal{worst}}^{\textnormal{d}} (\mathbf{y}/t;\delta)} = \overline{\gamma} \left( {\xi} t + \mathop{\sum}\limits_{\ell \in \mathcal{L}}{{\zeta'_\ell} {y_\ell}} + \mathop{\sum\sum}\limits_{1 \leq n < m \leq L} {{\mu_{n m}} \min(y_n,y_m)}  \right)  $}
\end{equation}
is concave in $(\mathbf{y},t)$. In the latter equation, we have exploited the fact that $\min(y_n/t,y_m/t) = \min(y_n,y_m)/t$, for any $t>0$. Regarding the perspective operation in the objective function, the following remark is helpful.

\begin{remark}[CVX implementation] \label{remark:CVX_implementation} 
In order to incorporate the objective function of problem \eqref{problem:Convex_relaxation} into the CVX package in Matlab \cite{CVX}, it is recommended to use MOSEK solver \cite{MOSEK} for guaranteed performance when using the $\log(\cdot)$ function. In this way, the successive approximation method, which is a heuristic approach, can be avoided. Specifically, we should utilize the relative-entropy function: $\operatorname{rel{\_}entr}(x,y) = x \log(x/y)$, which is \emph{convex} \cite{Boyd2004}. To be more precise, the (concave) objective function of \eqref{problem:Convex_relaxation} is expressed as 
\begin{equation}
\begin{split}
t \, \widehat{\operatorname{SE}}\vphantom{\widehat{\operatorname{SE}}}_{\textnormal{worst}}^\textnormal{d} (\mathbf{y}/t;\delta) &= t \log_2(1+{\widehat{\gamma}_{\textnormal{worst}}^\textnormal{d} (\mathbf{y}/t;\delta)})  \\
&= t \log_2( 1 + \overline{\gamma} \xi + u(\mathbf{y};\delta)/t)    \\
&= - \tfrac{1}{\log(2)} \operatorname{rel{\_}entr}(t,(1 + \overline{\gamma} \xi)t + u(\mathbf{y};\delta)) ,
\end{split}
\end{equation}
where 
\begin{equation}
u(\mathbf{y};\delta) = \overline{\gamma} \left( \sum_{\ell \in \mathcal{L}}{{\zeta'_\ell} {y_\ell}} + \mathop{\sum\sum}_{1 \leq n < m \leq L} {{\mu_{n m}} \min(y_n,y_m)} \right)
\end{equation}
is a concave function of $\mathbf{y}$. However, the arguments of relative entropy must be affine functions to be accepted by CVX. In order to overcome this obstacle, we introduce an auxiliary variable $w$ and an extra (convex) constraint $w \leq u(\mathbf{y};\delta)$. Ultimately, the convex problem that should be solved using CVX is formulated as  
\begin{subequations} \label{problem:Convex_relaxation_CVX}
\begin{alignat}{3}
  & \mathop {\textnormal{minimize}} \limits_{\mathbf{y},t,w} & \quad & \! \operatorname{rel{\_}entr}(t,(1 + \overline{\gamma} \xi)t + w)   \\
  & \,\textnormal{subject to} & & u(\mathbf{y};\delta) \geq (\gamma_{\min} - \overline{\gamma} \xi)t  ,  \\ 
  & & & {0 \leq y_\ell \leq t},\;\;\forall \ell \in \mathcal{L} ,  \\
  & & & t \, {\operatorname{P}_\textnormal{tot} (\mathbf{y}/t)} = 1  ,   \\
  & & & w \leq u(\mathbf{y};\delta)  .
\end{alignat}
\end{subequations}
This problem is equivalent to problem \eqref{problem:Convex_relaxation} because its optimum value is achieved when $w = u(\mathbf{y};\delta)$; observe that ${\partial \left( t \log_2( 1 + \overline{\gamma} \xi + w/t) \right)}/{\partial w} = \left[(1 + \overline{\gamma} \xi + w/t) \log(2)\right]^{-1} > 0$, and therefore $\operatorname{rel{\_}entr}(t,(1 + \overline{\gamma} \xi)t + w)$ is strictly decreasing with respect to $w$.  
\end{remark}

The proposed method to obtain an approximate solution to problem \eqref{problem:EE_d_original} is given in Algorithm \ref{algorithm:CRBM}. In steps 1-2, the convex relaxation problem \eqref{problem:Convex_relaxation}/\eqref{problem:Convex_relaxation_CVX} is solved, and then its optimal solution is used to compute a fractional (optimal) solution to problem \eqref{problem:Fractional_relaxation}. Subsequently, in steps 3-10, the fractional solution's entries are sorted in descending order, while initializing several programming variables. Next, based on that sorting, a successive activation of reflecting elements is performed to find the highest EE under the minimum-SNR constraint (steps 11-18). Finally, in steps 19-23, a feasible solution to the original problem \eqref{problem:EE_d_original} is reconstructed and returned, with its corresponding EE, by the algorithm. An approximation guarantee and the polynomial complexity of Algorithm \ref{algorithm:CRBM} are provided by the following theorem.

\begin{theorem} \label{theorem:CRBM}
Suppose that Assumptions \ref{assumption:CSI-uncertainty_radius} and \ref{assumption:minimum_quantization_bits} are true, and ${\gamma}_{\textnormal{worst}}^\textnormal{d} (\mathbf{1}_L;\delta) \geq \gamma_{\min}$. Then, the convex relaxation problem \eqref{problem:Convex_relaxation}/\eqref{problem:Convex_relaxation_CVX} is feasible, and Algorithm \ref{algorithm:CRBM} outputs a feasible solution $\widetilde{\mathbf{x}}$ to the original problem \eqref{problem:EE_d_original} such that $\widetilde{\operatorname{EE}} = {\operatorname{EE}_{\textnormal{worst}}^\textnormal{d}}(\widetilde{\mathbf{x}};\delta)$. In addition, the returned solution satisfies the following (a posteriori) performance guarantee\footnote{By saying \emph{``a posteriori''} we mean that the performance guarantee can be provided \emph{after} the termination of the algorithm.} 
\begin{equation} \label{equation:performance_guarantee}
0 \leq {\operatorname{EE}_{\textnormal{worst}}^{\textnormal{d},\star}} - \widetilde{\operatorname{EE}} \leq \operatorname{EE}_{\textnormal{rel}}^{\star} - \widetilde{\operatorname{EE}}  .
\end{equation}
Finally, the complexity of Algorithm \ref{algorithm:CRBM} is $O(L^{3.5})$, provided that the convex relaxation problem \eqref{problem:Convex_relaxation}/\eqref{problem:Convex_relaxation_CVX} is solved by an interior-point method. 
\end{theorem}

\begin{IEEEproof}
See Appendix \ref{appendix:CRBM}.   
\end{IEEEproof}

\noindent
\begin{minipage}[!t]{\columnwidth}
\begin{algorithm}[H]   
\caption{Convex-Relaxation-Based Method (CRBM)}  \label{algorithm:CRBM}
\small
\begin{algorithmic}[1] 
\State Solve the convex relaxation problem \eqref{problem:Convex_relaxation}/\eqref{problem:Convex_relaxation_CVX} to obtain a \newline (globally) optimal solution $(\mathbf{y}^\star,t^\star)$.
\State $\mathbf{x}^{\text{rel},\star} := {\mathbf{y}^\star}/{t^\star}$, where $\mathbf{x}^{\text{rel},\star} \in [0,1]^L$ (i.e., a fractional solution). 
\State Sort the entries of $\mathbf{x}^{\text{rel},\star} = [x_1^{\text{rel},\star},\dots,x_L^{\text{rel},\star}]^\top$ in descending order. Let  $\boldsymbol{\sigma} = [\sigma_1,\dots,\sigma_L]^\top$ be a permutation of $\mathcal{L}$ such that $x^{\text{rel},\star}_{\sigma_1} \geq \cdots \geq x^{\text{rel},\star}_{\sigma_L}$. 
\State $\Delta_{\Re} := \widehat{\alpha}_0$, $\Delta_{\Im} := 0$, $f_\text{d} :=  \widehat{\alpha}_0$, $\gamma^\text{d}_\text{worst} := \overline{\gamma} \left( f_\text{d} - \delta \right)^2$ 
\State $\operatorname{P}_\text{tot} :=  P_{\text{fix}} + L P_{\text{off}}$, $\operatorname{\Delta P} := P_{\text{on}} - P_{\text{off}}$ 
\If{$\gamma^\text{d}_\text{worst} \geq \gamma_{\min}$}
	\State  ${\operatorname{EE}_{\max}} := \log_2(1+\gamma^\text{d}_\text{worst})/\operatorname{P}_\text{tot}$, $M_{\max} := 0$ 
\Else 
	\State ${\operatorname{EE}_{\max}} := -\infty$ 
\EndIf 
\For{$M : = 1\ \text{to}\ L$}
	\State $\Delta_{\Re} := \Delta_{\Re} + {\widehat{\alpha}_{\sigma_M} \cos (\varepsilon_{\sigma_M})}$, $\Delta_{\Im} := \Delta_{\Im} + {\widehat{\alpha}_{\sigma_M} \sin (\varepsilon_{\sigma_M})}$ 
	\State $f_\text{d} := \sqrt{ \Delta_{\Re}^2 + \Delta_{\Im}^2 }$, $\gamma^\text{d}_\text{worst} := \overline{\gamma} \left( f_\text{d} - \delta \sqrt{1+M} \right)^2$ 
	\State $\operatorname{P}_\text{tot} :=  \operatorname{P}_\text{tot} + \operatorname{\Delta P}$, $\operatorname{EE} := \log_2(1+\gamma^\text{d}_\text{worst})/\operatorname{P}_\text{tot}$ 
	\If{$(\gamma^\text{d}_\text{worst} \geq \gamma_{\min}) \land (\operatorname{EE} > \operatorname{EE}_{\max})$}
		\State $\operatorname{EE}_{\max} := \operatorname{EE}$, $M_{\max} := M$    
	\EndIf	
\EndFor 
\State $\widetilde{\operatorname{EE}} := \operatorname{EE}_{\max}$, $\widetilde{\mathbf{x}} := \mathbf{0}_L$
\For{$m := 1\ \text{to}\ M_{\max}$}
	\State $\widetilde{x}_{\sigma_m} := 1$
\EndFor
\State \textbf{return} $(\widetilde{\operatorname{EE}},\widetilde{\mathbf{x}})$		
\end{algorithmic}
\end{algorithm} 
\end{minipage}

\begin{remark} 
The additional assumption that ${\gamma}_{\textnormal{worst}}^\textnormal{d} (\mathbf{1}_L;\delta) \geq \gamma_{\min}$ in Theorem \ref{theorem:CRBM} is relatively mild, because it is a necessary and sufficient condition for feasibility of the robust optimization problem \eqref{problem:EE_d_original}, provided that $b \geq 3$ (according to Proposition \ref{proposition:Feasibility_discrete}). In essence, its main purpose is to also include cases with $b = 2$. 
\end{remark}

\begin{remark}
The complexity of solving the convex relaxation problem \eqref{problem:Convex_relaxation}/\eqref{problem:Convex_relaxation_CVX}, using an interior-point method, is $O(L^{3.5})$: $O(\sqrt{L})$ Newton steps, each one with complexity $O(L^3)$. In practice, the number of Newton steps is (almost always) bounded by a constant, i.e., $O(1)$ \cite{Boyd2004}. As a consequence, the complexity of Algorithm \ref{algorithm:CRBM} is expected to be $O(L^3)$ for practical purposes. 
\end{remark}

\section{Numerical Results} \label{section:Numerical_Results}

In this section, we evaluate the performance and complexity of the proposed algorithms via simulations, where the transmitter, receiver and IRS are located at $(0,0,0)$, $(100,0,0)$ and $(50,20,10)$, respectively, with $(x,y,z)$-coordinates given in meters. All figures present average values derived from $100$ independent random realizations of channel estimates $\widehat{\mathbf{h}}$. In particular, $\widehat{h}_0 \sim \mathcal{CN}(0,\varrho_0)$, where $\varrho_0 = c_0^\text{ref} ({d_0}/{d_0^\text{ref}})^{-a_0}$ is the distance-dependent path loss, with $d_0$, $a_0$ being the Tx-Rx distance and path-loss exponent, and $d_0^\text{ref}$, $c_0^\text{ref}$ being the reference distance and path loss, respectively. Moreover, $\widehat{h}_\ell = {\beta_\ell} {\widehat{u}_\ell} {\widehat{v}_\ell}$, $\forall \ell \in \mathcal{L}$, where $\widehat{u}_\ell = \sqrt{\varrho_u} \left( \sqrt{\frac{\kappa_u}{1+\kappa_u}} {\widehat{u}_\ell^\text{LOS}} + \sqrt{\frac{1}{1+\kappa_u}} {\widehat{u}_\ell^\text{NLOS}} \right)$ and $\widehat{v}_\ell =  \sqrt{\varrho_v} \left( \sqrt{\frac{\kappa_v}{1+\kappa_v}} {\widehat{v}_\ell^\text{LOS}} + \sqrt{\frac{1}{1+\kappa_v}} {\widehat{v}_\ell^\text{NLOS}} \right)$, with $\kappa_{u/v}$ being the Rician factors and $\varrho_{u/v} = c_{u/v}^\text{ref} ({d_{u/v}}/{d_{u/v}^\text{ref}})^{-a_{u/v}}$; the subscripts $u$ and $v$ refer to Tx-IRS and IRS-Rx links, respectively. Assuming a uniform linear array of reflecting elements at the IRS (parallel to the $x$-axis), the line-of-sight (LOS) components are given by $\widehat{u}_\ell^\text{LOS} = e^{j2\pi \tfrac{d}{\lambda} (\ell - 1) \sin(\vartheta^\text{A}) \cos(\varphi^\text{A})}$ and $\widehat{v}_\ell^\text{LOS} = e^{j2\pi \tfrac{d}{\lambda} (\ell - 1) \sin(\vartheta^\text{D}) \cos(\varphi^\text{D})}$, where $d$ is the distance between adjacent IRS elements, and $\lambda$ is the wavelength of the carrier frequency. In addition, $\vartheta^\text{A}$ and $\varphi^\text{A}$ are the inclination and azimuth angle-of-arrival (AoA) of signals from the Tx to the IRS, respectively. Similarly, $\vartheta^\text{D}$ and $\varphi^\text{D}$ are the inclination and azimuth angle-of-departure (AoD) of signals from the IRS to the Rx, respectively \cite{Wang2019}. The non-line-of-sight (NLOS) components ${\widehat{u}_\ell^\text{NLOS}}, {\widehat{v}_\ell^\text{NLOS}} \sim \mathcal{CN}(0,1)$. The simulation parameters are $c_0^\text{ref} = 10^{-5}$, $c_u^\text{ref} = c_v^\text{ref} = 10^{-3}$, $d_0^\text{ref} = d_u^\text{ref} = d_v^\text{ref} = 1\ \text{m}$, $a_0 = 3.7$, $a_u = a_v = 2.2$, $\kappa_u = \kappa_v = 5\ \text{dB}$, and $d/\lambda = 0.5$.

\begin{figure*}[!t]
\centering
\begin{subfigure}{.5\textwidth}
  \centering
  \includegraphics[width=\linewidth]{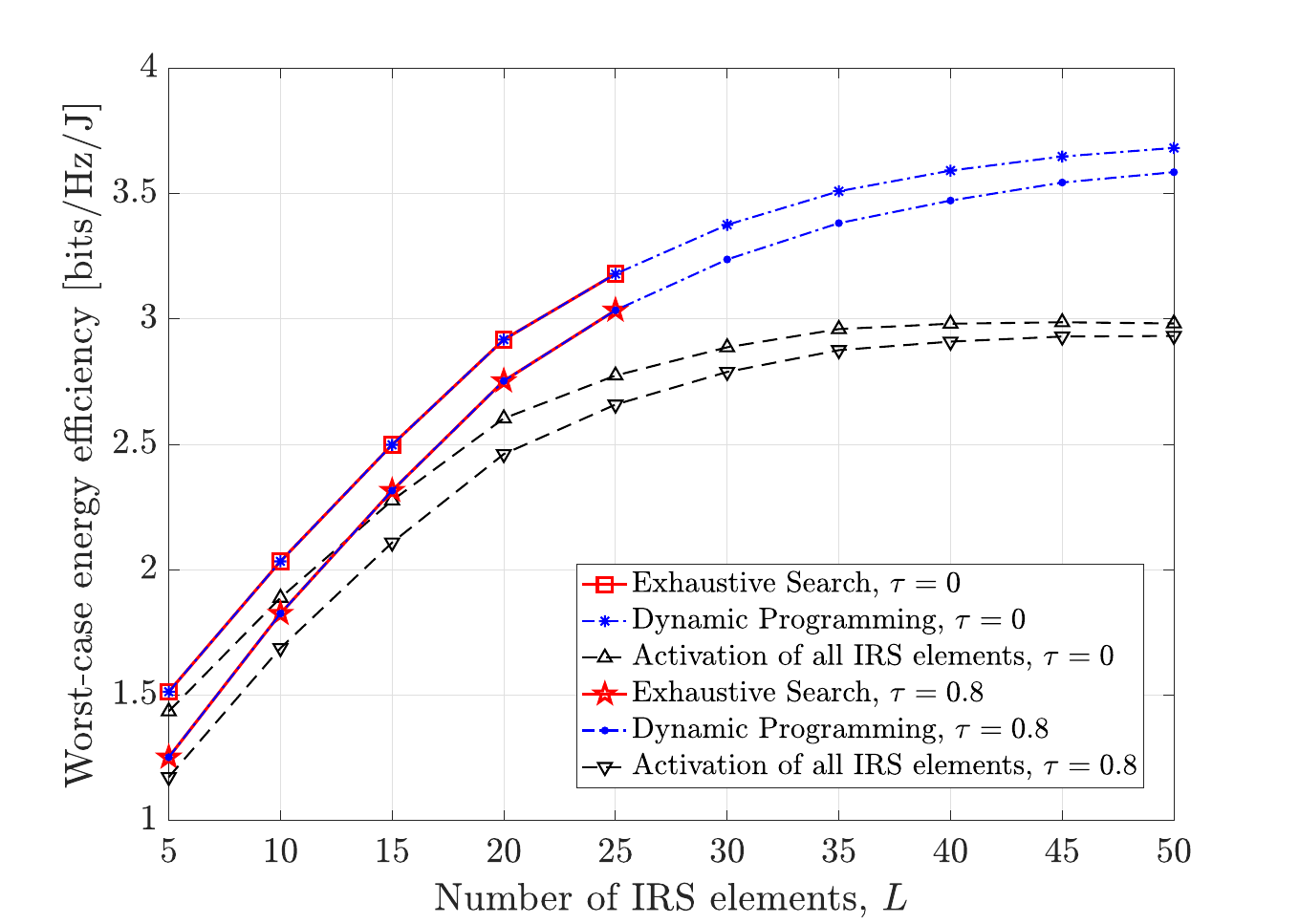}
  \caption{Continuous phase shifts}
  \label{figure:EE_vs_L_continuous}
\end{subfigure}%
\begin{subfigure}{.5\textwidth}
  \centering
  \includegraphics[width=\linewidth]{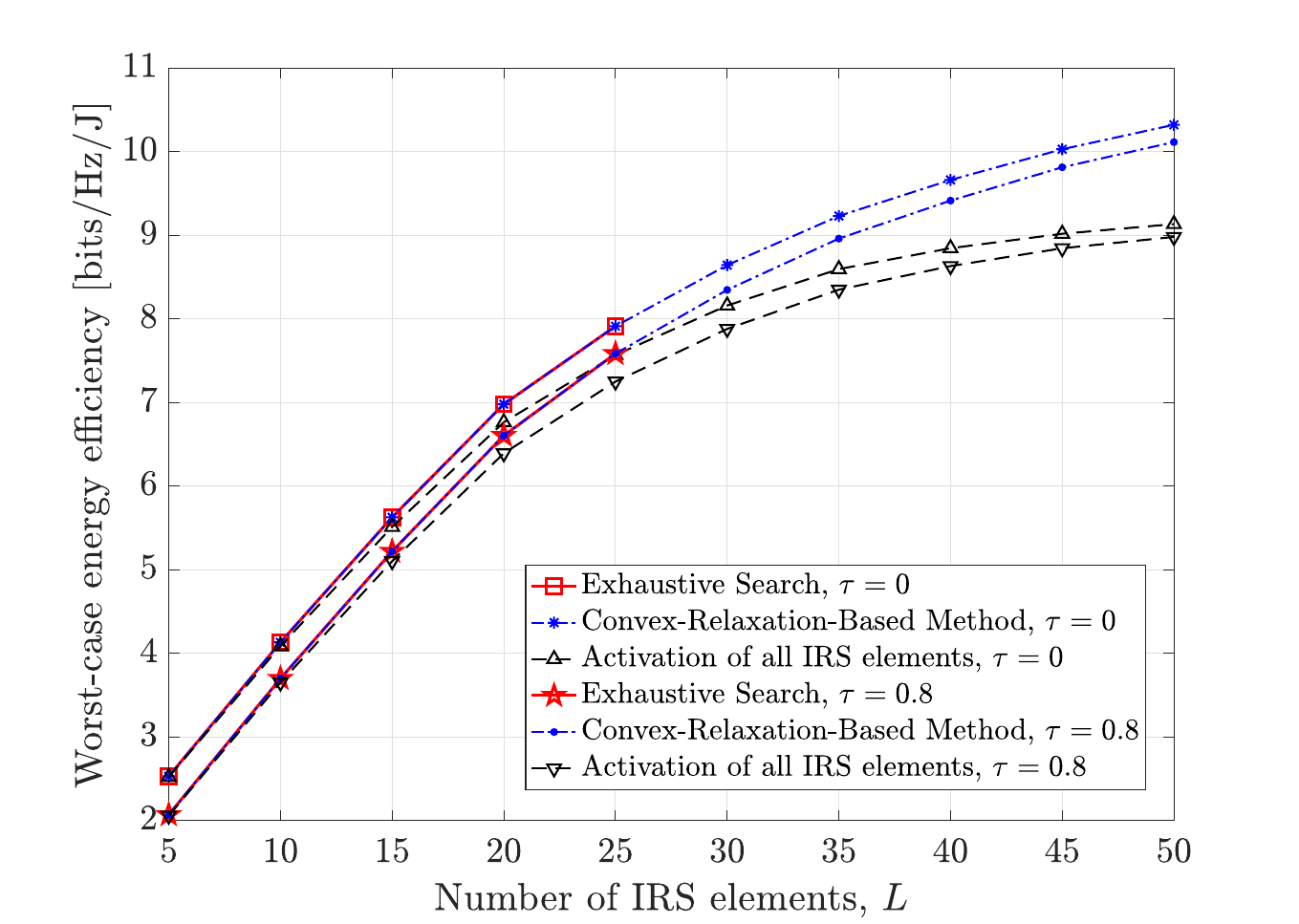}
  \caption{Discrete phase shifts}
  \label{figure:EE_vs_L_discrete}
\end{subfigure}
\caption{Worst-case energy efficiency versus the number of IRS elements, with CSI-uncertainty radius $\delta = \tau {\widehat{\alpha}_{\min}}$.}
\label{figure:EE_vs_L}
\end{figure*}

\begin{figure*}[!t]
\centering
\begin{subfigure}{.5\textwidth}
  \centering
  \includegraphics[width=\linewidth]{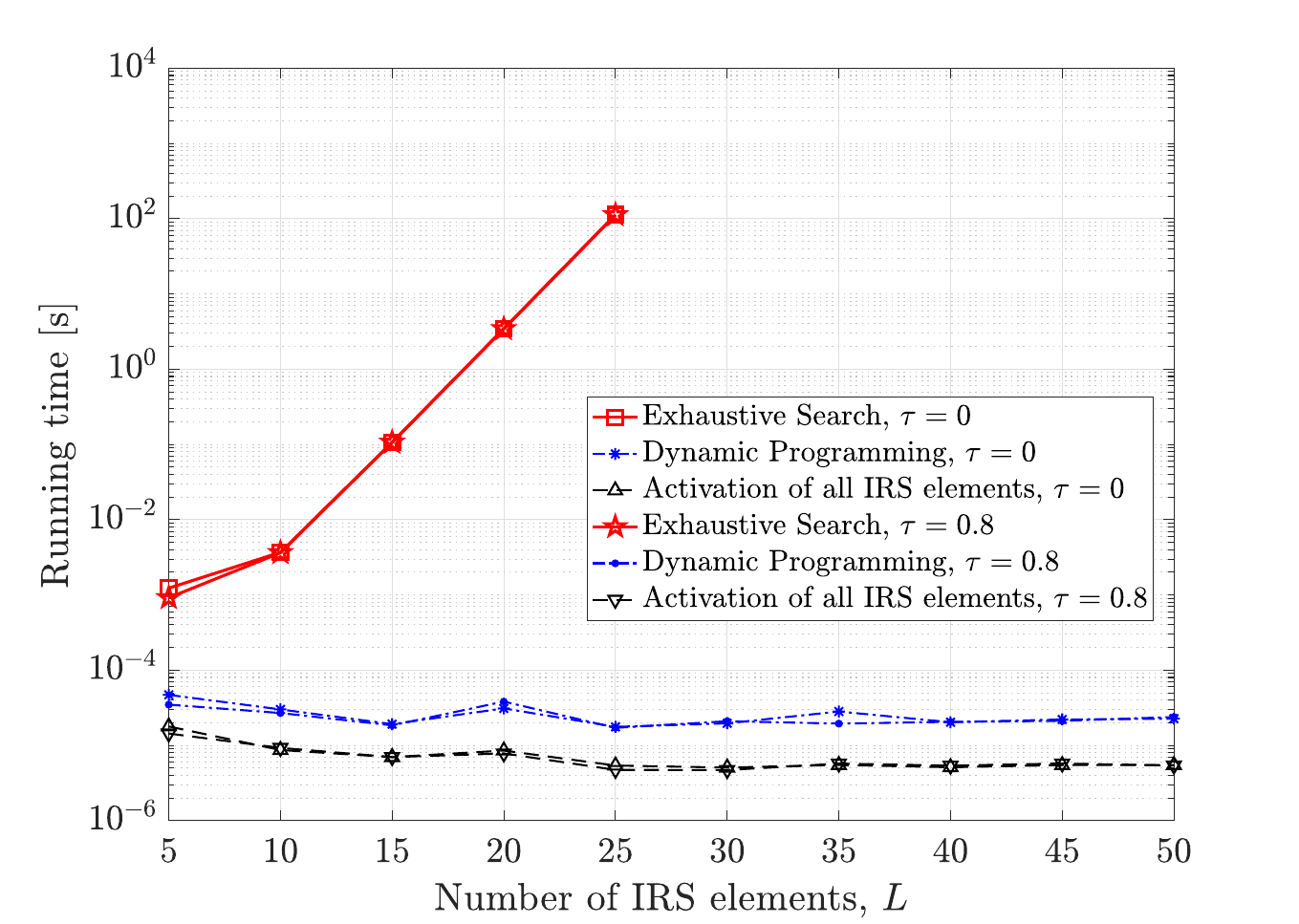}
  \caption{Continuous phase shifts}
  \label{figure:CPU_time_vs_L_continuous}
\end{subfigure}%
\begin{subfigure}{.5\textwidth}
  \centering
  \includegraphics[width=\linewidth]{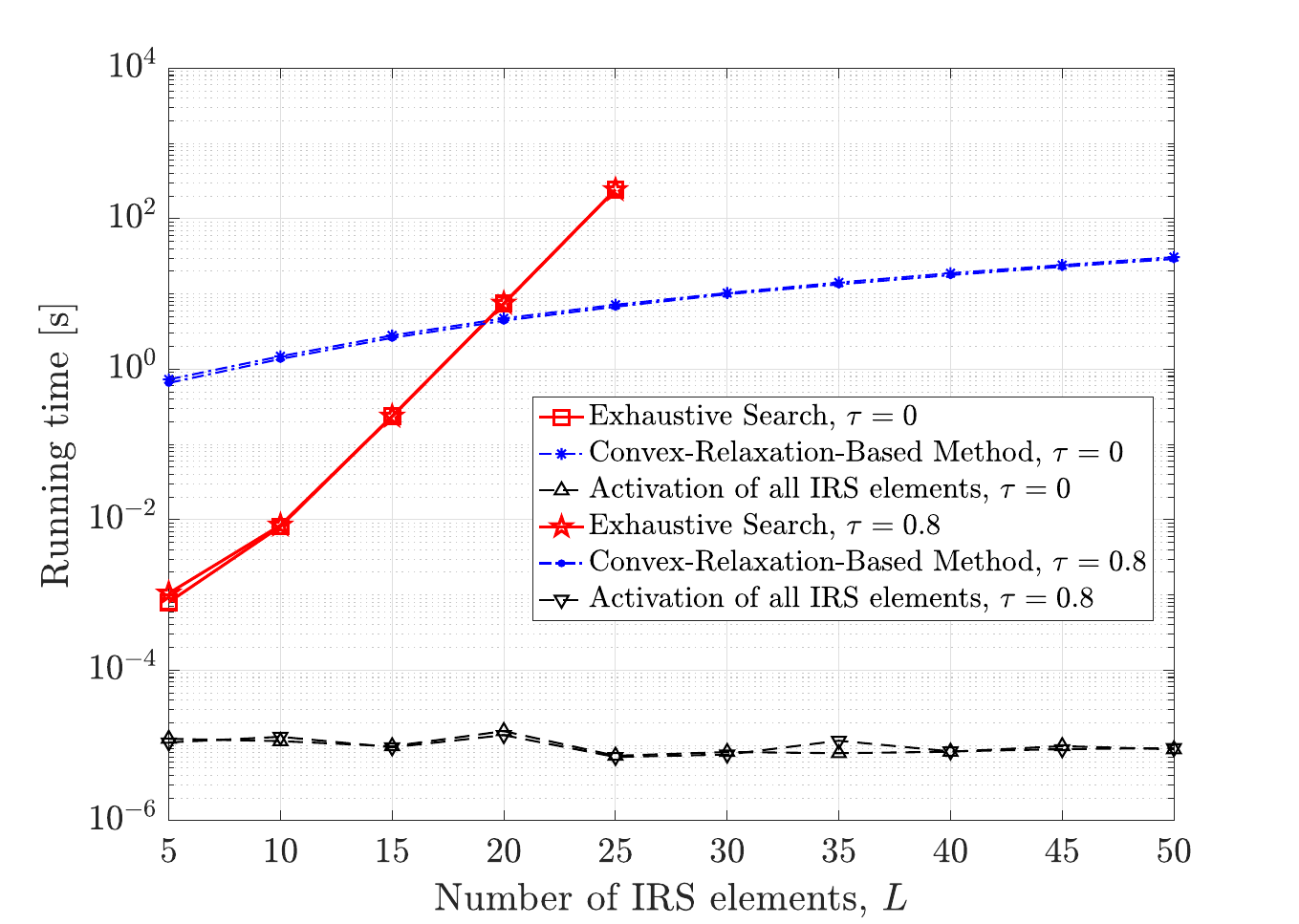}
  \caption{Discrete phase shifts}
  \label{figure:CPU_time_vs_L_discrete}
\end{subfigure}
\caption{Running time versus the number of IRS elements, with CSI-uncertainty radius $\delta = \tau {\widehat{\alpha}_{\min}}$, using a digital computer with Intel Core i5-2400 CPU (3.1 GHz) and 8 GB RAM.} 
\label{figure:CPU_time_vs_L}
\end{figure*}

The CSI-uncertainty radius is selected as $\delta = \tau {\widehat{\alpha}_{\min}}$, where $\tau \in [0,1]$ so that $\delta \leq {\widehat{\alpha}_{\min}}$, thus satisfying Assumption \ref{assumption:CSI-uncertainty_radius}. The ideal case without channel estimation errors (i.e., \emph{perfect CSI}) is obtained when $\tau = 0$. The minimum SNR is chosen to be $\gamma^{\text{d/c}}_{\min} = {\nu} {\gamma}^{\text{d/c}}_{\text{worst}} (\mathbf{1}_L;\widehat{\alpha}_{\min})$, where $\nu \in [0,1]$ is the minimum-SNR control parameter. In this way, the feasibility of the generated optimization problems is guaranteed, because ${\gamma}^{\text{d/c}}_{\text{worst}} (\mathbf{1}_L;\delta) \geq {\gamma}^{\text{d/c}}_{\text{worst}} (\mathbf{1}_L;\widehat{\alpha}_{\min}) \geq \gamma^{\text{d/c}}_{\min}$ (see Propositions \ref{proposition:Optimal_worst-case_EE_monotonicity_discrete} and \ref{proposition:Optimal_worst-case_EE_monotonicity_continuous}). Unless otherwise stated, the remaining simulation parameters are the following: $L = 20$,\footnote{We have used a medium number of IRS elements in order to be able to run the exhaustive-search algorithm required for comparison purposes. Nevertheless, Figs. \ref{figure:EE_vs_L} and \ref{figure:CPU_time_vs_L} show numerical results for large $L$ as well.} $b = 4$ (for discrete phase shifts), $p = 15\ \text{dBm}$, $\sigma_w^2 = -95\ \text{dBm}$, $\beta_\ell = 0.9,\ \forall \ell \in \mathcal{L}$, $\eta = 0.8$, $P_{\text{static}} = 10\ \text{mW}$, $P_\text{on} = 15\ \text{mW}$ (for continuous phase shifts), $P_{\text{on}}(b) = 1.8 b - 3$ in mW (for discrete phase shifts with $b \geq 2$),\footnote{Similar values for the power consumption of each phase shifter were also used in \cite{Huang2019}.} $P_\text{off} = 0.3\ \text{mW}$, and $\nu = 0.7$. In addition, we have used CVX \cite{CVX} with MOSEK \cite{MOSEK} to solve the convex relaxation problem \eqref{problem:Convex_relaxation}/\eqref{problem:Convex_relaxation_CVX} in Algorithm \ref{algorithm:CRBM}  (see Remark \ref{remark:CVX_implementation}). Besides the exhaustive-search method that attains the global maximum, we also consider a \emph{baseline scheme}, namely, the \emph{activation of all IRS elements} (i.e., $\mathbf{x} = \mathbf{1}_L$).\footnote{Note that $\mathbf{x} = \mathbf{1}_L$ is an optimal solution to the worst-case SE maximization, under the minimum-SNR constraint, as long as $b \geq 3$. This is because Assumption \ref{assumption:CSI-uncertainty_radius} is true in all simulation scenarios, and ${\gamma}^{\text{d/c}}_{\textnormal{worst}} (\mathbf{1}_L;\delta) \geq {\gamma^{\text{d/c}}_{\text{worst}}} (\mathbf{x};\delta)$, $\forall \mathbf{x} \in \{0,1\}^L$, due to Propositions \ref{proposition:Feasibility_discrete} and \ref{proposition:Feasibility_continuous}.}

First of all, Fig. \ref{figure:EE_vs_L} shows the worst-case EE versus the number of IRS elements for continuous and discrete phase shifts.\footnote{We have run the exhaustive-search algorithm only for $L \leq 25$, because of its exponential complexity.} For each algorithm, the worst-case EE decreases as the CSI-uncertainty radius increases, which is in agreement with Propositions \ref{proposition:Optimal_worst-case_EE_monotonicity_discrete} and \ref{proposition:Optimal_worst-case_EE_monotonicity_continuous} (note that $\gamma^{\text{d/c}}_{\min}$ is independent of $\delta$). Consequently, the highest worst-case EE occurs in the perfect-CSI regime, i.e., for $\delta = \tau = 0$. In Fig. \ref{figure:EE_vs_L_continuous}, the proposed DP algorithm has identical performance with the exhaustive search (as expected by Theorem \ref{theorem:DP_algorithm}), while performing much better than the baseline scheme, especially for large $L$. In Fig. \ref{figure:EE_vs_L_discrete}, CRBM shows very similar performance as the exhaustive search, whereas the performance gap between CRBM and the baseline scheme increases with the number of IRS elements. For small $L$, however, all algorithms achieve roughly the same worst-case EE; this is because $P_{\text{fix}} + L P_{\text{off}} \gg (P_{\text{on}} - P_{\text{off}}) {\sum_{\ell \in \mathcal{L}} {x_\ell}} \implies \operatorname{P}_\text{tot} (\mathbf{x}) \approx P_{\text{fix}} + L P_{\text{off}}, \ \forall \mathbf{x} \in \{0,1\}^L$, and therefore EE maximization approximately reduces to SE maximization (i.e., the activation of all IRS elements). We can also observe that, for $L = 50$, DP and CRBM achieve approximately $23\%$ and $13\%$ EE gain, respectively, in comparison with the activation of all IRS elements.

In addition, Fig. \ref{figure:CPU_time_vs_L} illustrates the running time of algorithms as a function of $L$, where the baseline scheme exhibits the lowest complexity. In Fig. \ref{figure:CPU_time_vs_L_continuous}, the DP's runtime is of the order of $10^{-5}\ \text{s}$ for all $L$, whereas that of exhaustive search grows rapidly with $L$. For example, DP is about \emph{seven orders of magnitude faster} than the exhaustive search when $L=25$. According to Fig. \ref{figure:CPU_time_vs_L_discrete}, the proposed CRBM is slower compared to the exhaustive search for small $L$. Nevertheless, CRBM is expected to be \emph{many orders of magnitude faster} than the exhaustive search for large $L$, because the complexity of the latter algorithm increases exponentially with $L$. Overall, the running times of DP and CRBM scale mildly with the number of IRS elements, thus achieving \emph{remarkable trade-offs between performance and complexity}.

\begin{figure*}[!t]
\centering
\begin{subfigure}{.5\textwidth}
  \centering
  \includegraphics[width=\linewidth]{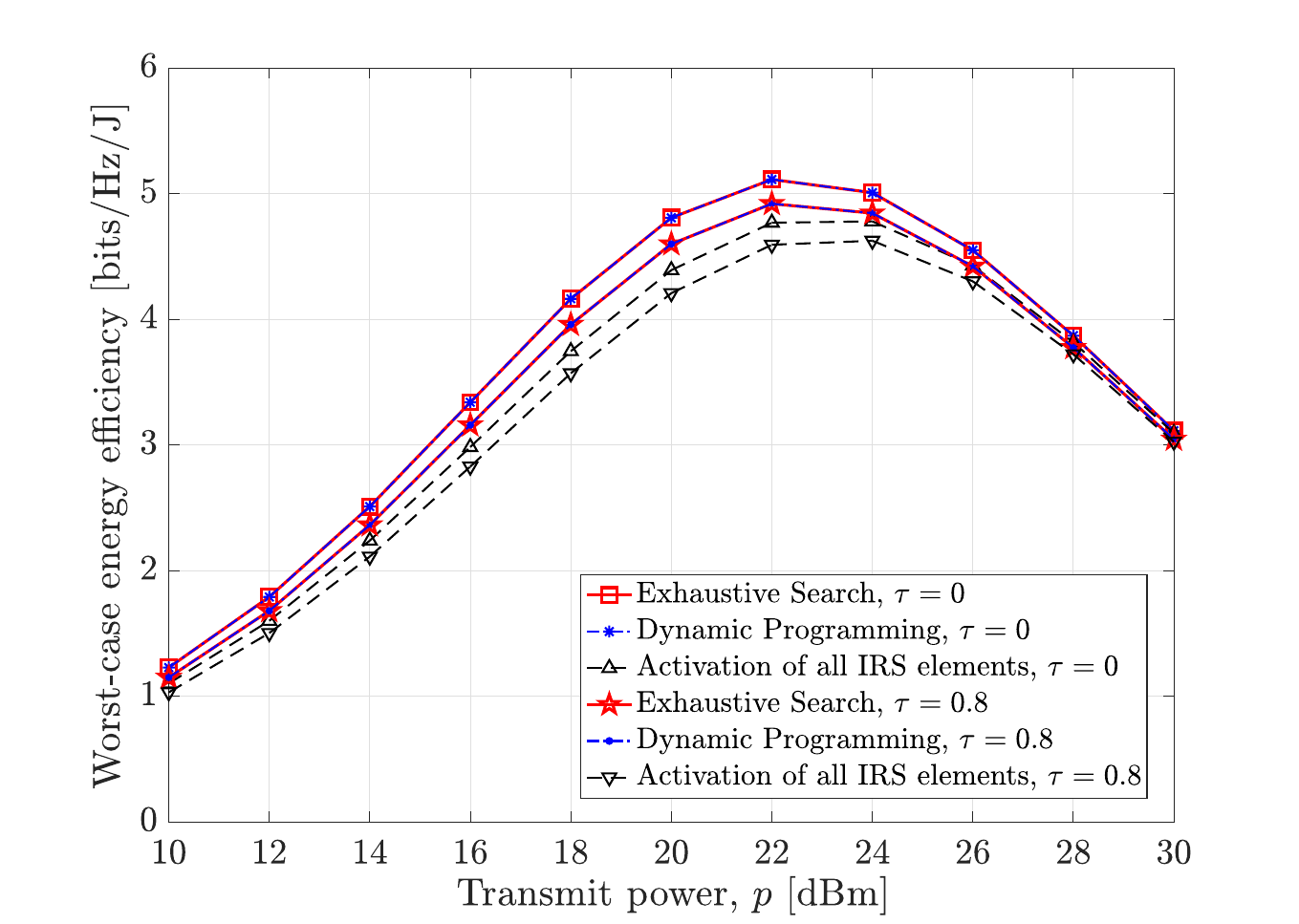}
  \caption{Continuous phase shifts}
  \label{figure:EE_vs_p_continuous}
\end{subfigure}%
\begin{subfigure}{.5\textwidth}
  \centering
  \includegraphics[width=\linewidth]{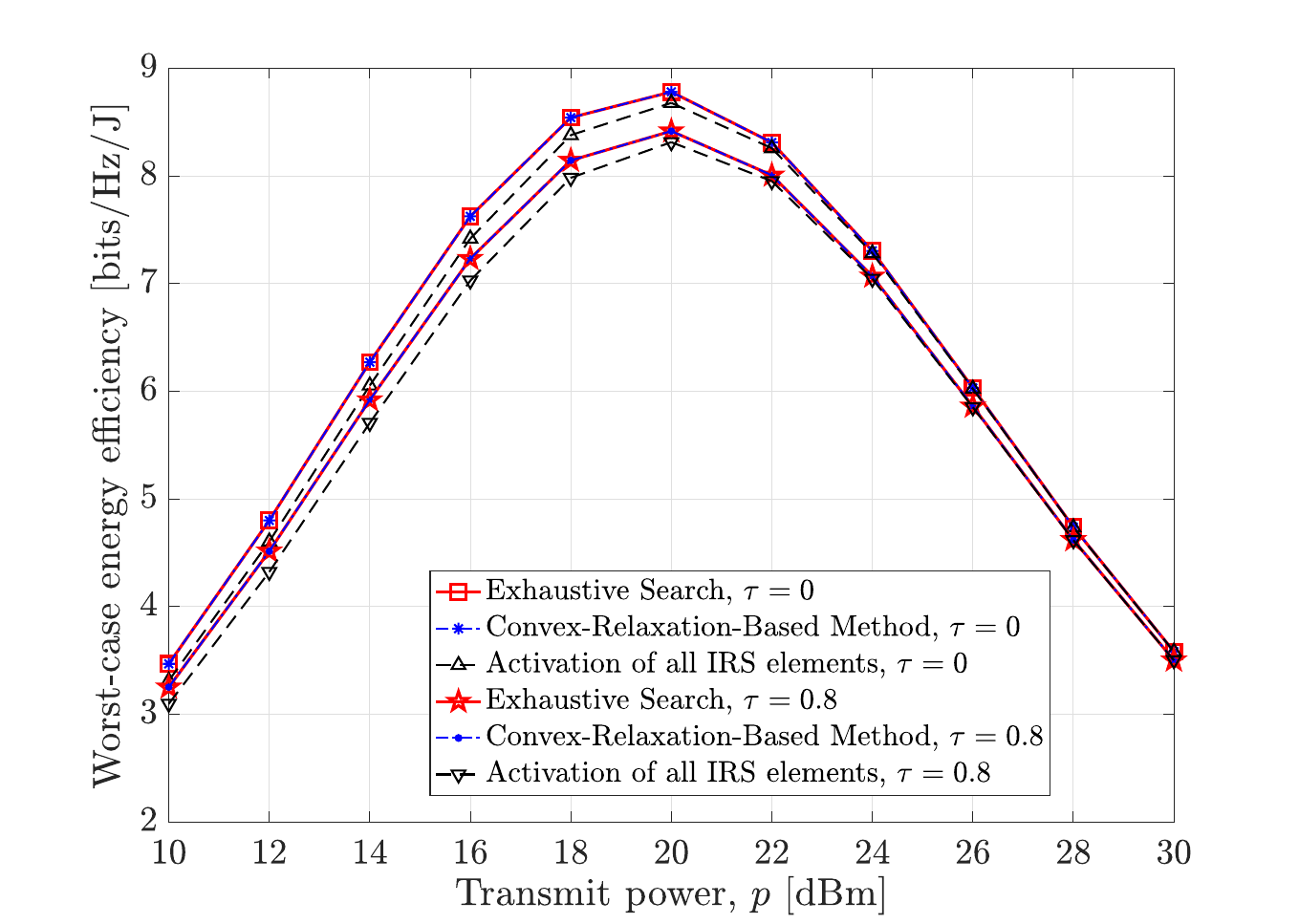}
  \caption{Discrete phase shifts}
  \label{figure:EE_vs_p_discrete}
\end{subfigure}
\caption{Worst-case energy efficiency versus the transmit power, with CSI-uncertainty radius $\delta = \tau {\widehat{\alpha}_{\min}}$.}
\label{figure:EE_vs_p}
\end{figure*}

\begin{figure*}[!t]
\centering
\begin{subfigure}{.5\textwidth}
  \centering
  \includegraphics[width=\linewidth]{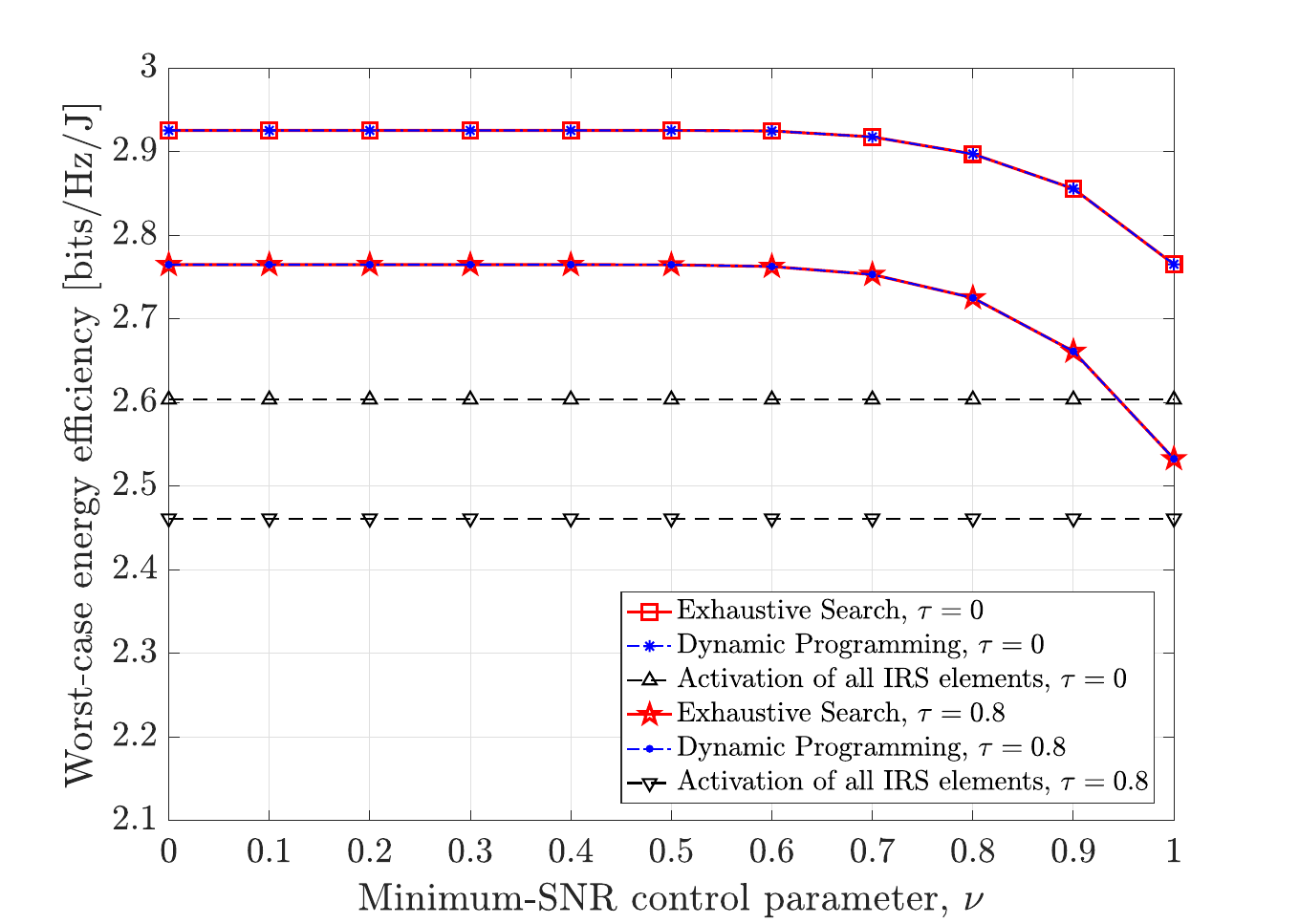}
  \caption{Continuous phase shifts}
  \label{figure:EE_vs_nu_continuous}
\end{subfigure}%
\begin{subfigure}{.5\textwidth}
  \centering
  \includegraphics[width=\linewidth]{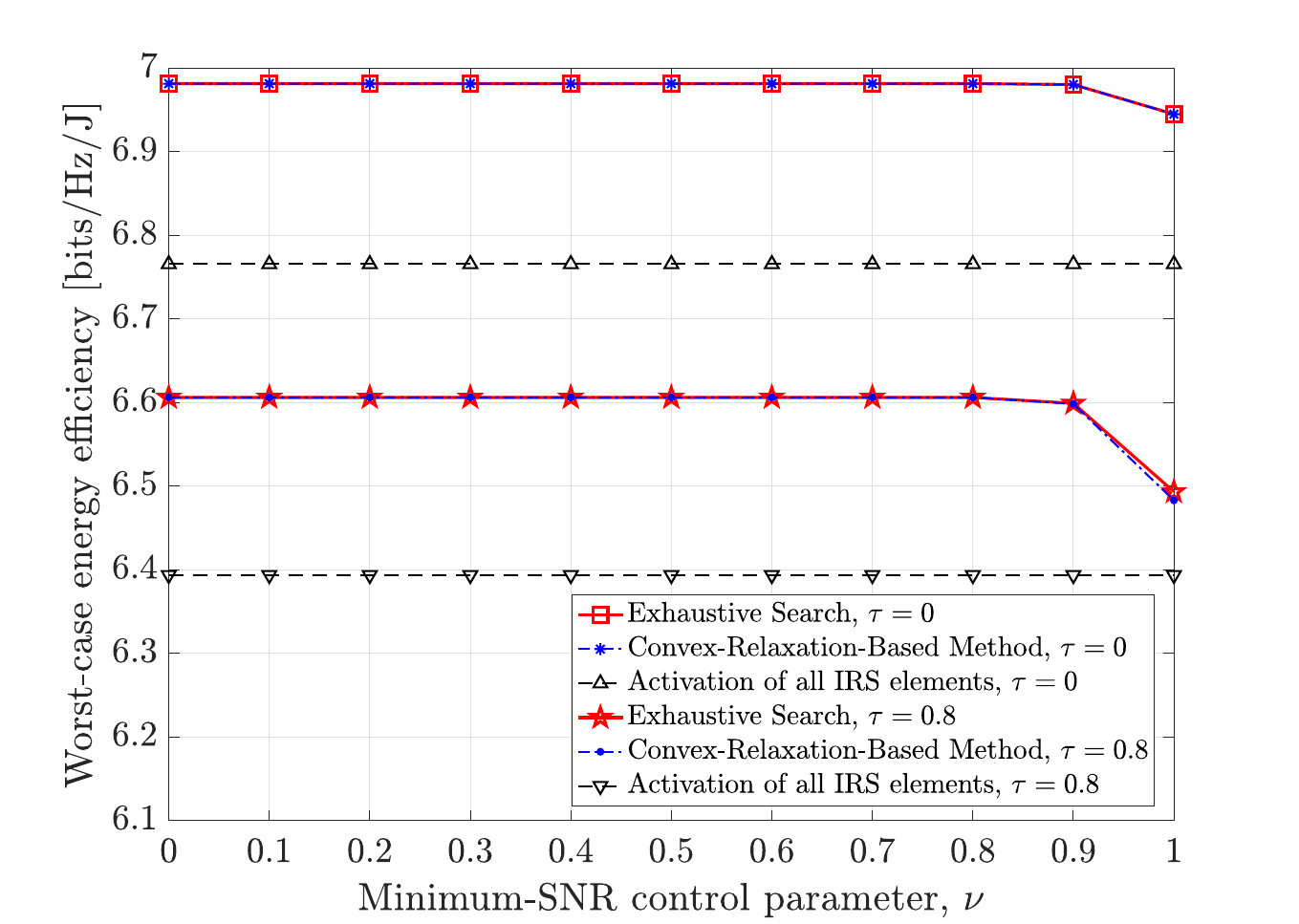}
  \caption{Discrete phase shifts}
  \label{figure:EE_vs_nu_discrete}
\end{subfigure}
\caption{Worst-case energy efficiency versus the minimum-SNR control parameter, with $\gamma^{\text{d/c}}_{\min} = {\nu} {\gamma}^{\text{d/c}}_{\text{worst}} (\mathbf{1}_L;\widehat{\alpha}_{\min})$ and CSI-uncertainty radius $\delta = \tau {\widehat{\alpha}_{\min}}$.}
\label{figure:EE_vs_nu}
\end{figure*}

Moreover, Fig. \ref{figure:EE_vs_p} presents the worst-case EE against the transmit power. Again, the performance of DP/CRBM coincides with that of the exhaustive search, and is higher than that of the benchmark. We can also observe the monotonicity of the worst-case EE with respect to the CSI-uncertainty radius, in line with Propositions \ref{proposition:Optimal_worst-case_EE_monotonicity_discrete} and \ref{proposition:Optimal_worst-case_EE_monotonicity_continuous}. All algorithms attain their maximum for some value of $p$, while they show similar performance for high transmit power (in this case, $\operatorname{P}_\text{tot} (\mathbf{x}) \approx P_{\text{fix}} + L P_{\text{off}} \approx p/\eta, \ \forall \mathbf{x} \in \{0,1\}^L$, thus EE maximization tends to be equivalent to SE maximization as $p \to \infty$).

Furthermore, we examine the impact of the minimum required SNR on the worst-case EE. Based on Fig. \ref{figure:EE_vs_nu}, we can make similar observations as in the previous figures. In addition, the worst-case EE achieved by DP, CRBM and exhaustive search is a nonincreasing function of $\nu$, since the increase of $\nu$ (or $\gamma^{\text{d/c}}_{\min}$) results in the shrinkage/contraction of the feasible sets of problems \eqref{problem:EE_d_original} and \eqref{problem:EE_c_original}; see also Propositions \ref{proposition:Optimal_worst-case_EE_monotonicity_discrete} and \ref{proposition:Optimal_worst-case_EE_monotonicity_continuous}.

\begin{figure}[!t]
\centering
\includegraphics[width=\linewidth]{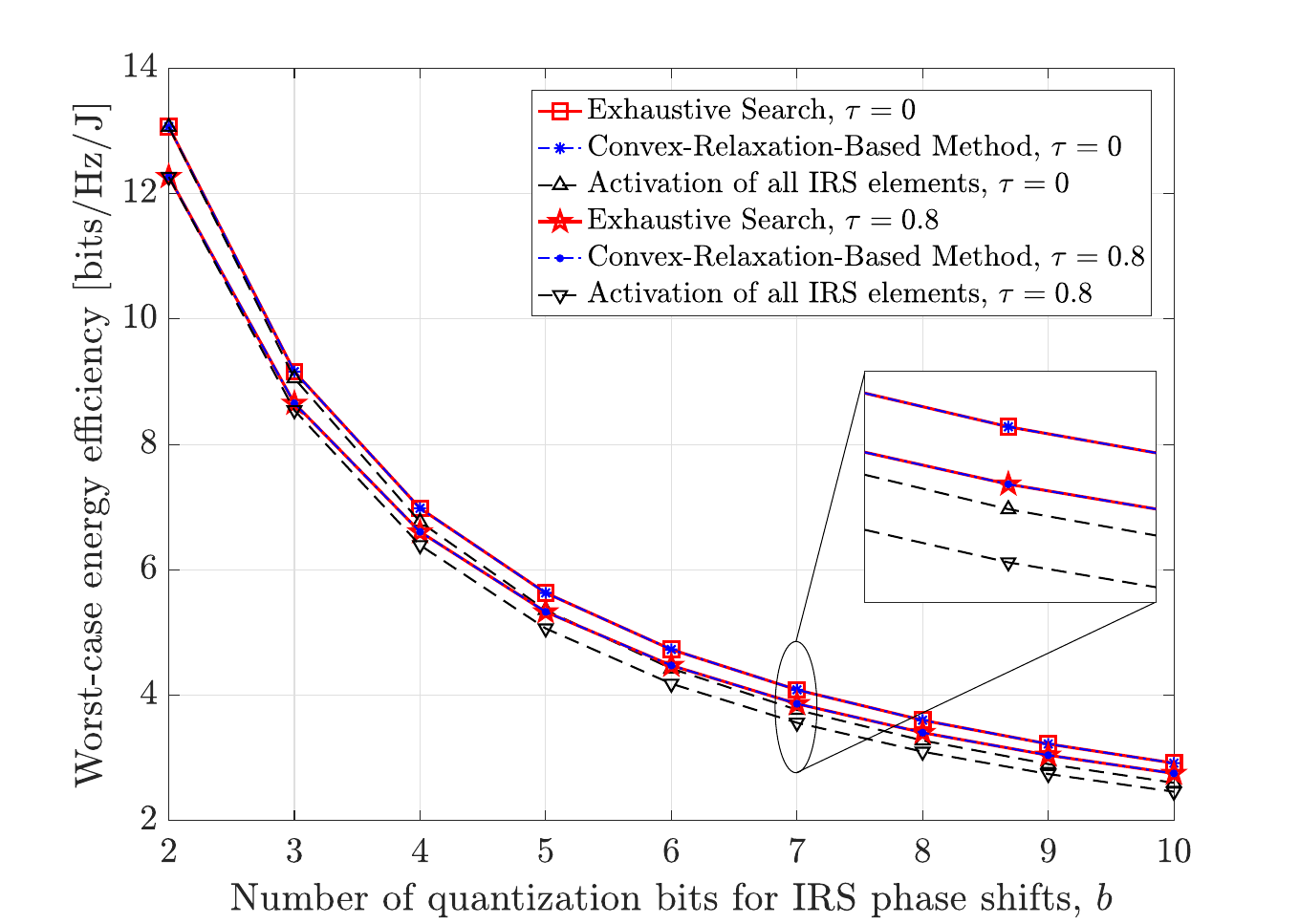}
\caption{Worst-case energy efficiency versus the number of quantization bits for (discrete) IRS phase shifts, with CSI-uncertainty radius $\delta = \tau {\widehat{\alpha}_{\min}}$.}
\label{figure:EE_vs_b}
\end{figure}

Finally, the effect of the number of quantization bits for discrete IRS phase shifts on the system performance is shown in Fig. \ref{figure:EE_vs_b}. For all algorithms, the worst-case EE decreases as $b$ increases, since larger $b$ leads to higher power consumption per activated IRS element. It is also interesting to observe that CRBM achieves the global maximum, while outperforming the baseline method for both values of $\tau$.

\section{Conclusion and Future Work} \label{section:Conclusion}

In this paper, we dealt with the maximization of the worst-case EE in IRS-assisted communication systems with imperfect CSI, via robust activation of IRS elements. In particular, we developed two polynomial-time algorithms based on dynamic programming and convex relaxation for continuous and discrete phase shifts, respectively. Due to their low complexity, these algorithms are suitable for solving large-scale problems, i.e., with large number of IRS elements. Furthermore, numerical results verified the effectiveness of the proposed methods in comparison with the exhaustive search and a conventional scheme. 

An interesting direction for future research is the joint power allocation and IRS-element activation, which is a mixed-integer optimization problem. Moreover, the extension to multi-user multiple-input multiple-output (MIMO) systems remains an open and challenging research topic.

\appendices

\section{Proof of Theorem \ref{theorem:discrete_phase-shifts_closed-form}} \label{appendix:discrete_phase-shifts_closed-form}

Concerning the equivalence between \eqref{equation:discrete_phase-shifts} and \eqref{equation:discrete_phase-shifts_closed-form}, it is sufficient to show that, given any $\phi_\ell^\star \in [0,2\pi)$,
\begin{equation}
\begin{split}
& \lambda_\ell \in \{ 0,\dots,K-1 \} \ \land \ \phi_\ell^\star \in \mathcal{R}_{\lambda_\ell} \\  
& \!\! \iff  \lambda_\ell = \operatorname{round}({\phi_\ell^\star} / \omega) \bmod K .
\end{split}
\end{equation}

In the forward direction ($\Longrightarrow$), we examine two cases: i) $\lambda_\ell = 0$ and ii) $\lambda_\ell = k$, where $k \in \{ 1,\dots,K-1 \}$. In the first case, either $\phi_\ell^\star \in [0,\omega/2) \implies \operatorname{round}({\phi_\ell^\star} / \omega) = 0$, or $\phi_\ell^\star \in [2\pi - \omega/2, 2\pi) \implies \operatorname{round}({\phi_\ell^\star} / \omega) = K$. In both sub-cases we have $\operatorname{round}({\phi_\ell^\star} / \omega) \bmod K = 0 = \lambda_\ell$. In the second case, $\phi_\ell^\star \in [k\omega - \omega/2, k\omega + \omega/2) \implies \operatorname{round}({\phi_\ell^\star} / \omega) = k$, thus $\operatorname{round}({\phi_\ell^\star} / \omega) \bmod K = k = \lambda_\ell$.

In the backward direction ($\Longleftarrow$), obviously $ \lambda_\ell \in \{ 0,\dots,K-1 \}$. Note that $\phi_\ell^\star \in [0,2\pi) \implies {{\phi_\ell^\star} / \omega} \in [0,K)$, and thus $\operatorname{round}({\phi_\ell^\star} / \omega) \in \{0,\dots,K \}$. We examine again two cases: i) $\lambda_\ell = 0$ and ii) $\lambda_\ell = k$, where $k \in \{ 1,\dots,K-1 \}$. In the first case, either $\operatorname{round}({\phi_\ell^\star} / \omega) = 0 \implies {\phi_\ell^\star} / \omega + 1/2 - 1 < 0 \leq {\phi_\ell^\star} / \omega + 1/2 \implies 0 = \max (0,-1/2) \leq {\phi_\ell^\star} / \omega < 1/2$,  or $\operatorname{round}({\phi_\ell^\star} / \omega) = K \implies {\phi_\ell^\star} / \omega + 1/2 - 1 < K \leq {\phi_\ell^\star} / \omega + 1/2 \implies K - 1/2 \leq {\phi_\ell^\star} / \omega < \min(K,K + 1/2) = K$. Therefore, $\phi_\ell^\star \in \mathcal{R}_0 = \mathcal{R}_{\lambda_\ell}$. In the second case, $\operatorname{round}({\phi_\ell^\star} / \omega) = k \implies {\phi_\ell^\star} / \omega + 1/2 - 1 < k \leq {\phi_\ell^\star} / \omega + 1/2 \implies k - 1/2 \leq {\phi_\ell^\star} / \omega < k + 1/2 $. As a result, $\phi_\ell^\star \in \mathcal{R}_k = \mathcal{R}_{\lambda_\ell}$. 

Furthermore, the total complexity of \eqref{equation:discrete_phase-shifts_closed-form} is $O(L \log K)$, because the modulo operation has complexity $O(\log K)$, division and multiplication by $K$, and is computed $L$ times.

\section{Proof of Theorem \ref{theorem:Worst-case_SNR_discrete}} \label{appendix:Worst-case_SNR_discrete}

Firstly, observe that constraint \eqref{constraint:CSI_uncertainty} does not impose any restrictions on the principal arguments of $\widetilde{\mathbf{h}}$, apart from being in the interval $[0,2\pi)$, but only on its magnitudes. Consequently, by inspection of \eqref{equation:SNR_discrete_phase_shifts}, we can deduce that the minimum SNR is achieved when  
\begin{equation}
\widetilde{\theta}_0^\star = (\vartheta(\mathbf{x}) + \widehat{\theta}_0 + \pi) \bmod {2\pi}  ,
\end{equation}
\begin{equation}
\widetilde{\theta}_\ell^\star = (\vartheta(\mathbf{x}) + \widehat{\theta}_0 - {\phi}_\ell^{\text{d}} + \pi) \bmod {2\pi} ,  \;\;\forall \ell \in \mathcal{L} .
\end{equation}
In other words, the arguments of $\widetilde{\mathbf{h}}$ are adjusted so as to cause the largest possible decrease in $f_\text{d}(\mathbf{x})$; in particular, $\widetilde{h}_0 + \sum_{\ell \in \mathcal{L}} {x_\ell \widetilde{h}_\ell e^{j{\phi}_\ell^{\text{d}}}}$ is aligned with $f_\text{d}(\mathbf{x})  e^{j(\vartheta(\mathbf{x})+\widehat{\theta}_0)}$, but in the opposite direction. 

Hence, problem \eqref{problem:Worst-case_SNR_discrete} is equivalent to the following problem 
\begin{subequations} \label{problem:Absolute_value}
\begin{alignat}{3}
  & \mathop {\text{minimize}} \limits_{\widetilde{\boldsymbol{\alpha}} \in \mathbb{R}_{+}^{L+1}} & \quad &  \left| f_\text{d}(\mathbf{x}) - \left( \widetilde{\alpha}_0 + \sum_{\ell \in \mathcal{L}} {x_\ell \widetilde{\alpha}_\ell} \right) \right|   \\
  & \,\text{subject to} & & \sqrt{ \widetilde{\alpha}_0^2 + \sum_{\ell \in \mathcal{L}} {\widetilde{\alpha}_\ell^2} } \leq \delta ,
\end{alignat}
\end{subequations}
where $\widetilde{\boldsymbol{\alpha}} = [{\widetilde{\alpha}_0}, \ldots ,{\widetilde{\alpha}_L}]^\top$. In order to facilitate the solution of problem \eqref{problem:Absolute_value}, we exploit the following lemma.

\begin{lemma}
Let $g(\mathbf{x};\delta)$ be the global maximum of the following optimization problem
\begin{subequations}  \label{problem:g}
\begin{alignat}{3}
 g(\mathbf{x};\delta) \! \triangleq & \mathop {\textnormal{maximize}} \limits_{\widetilde{\boldsymbol{\alpha}} \in \mathbb{R}_{+}^{L+1}} & \quad & \widetilde{\alpha}_0 + \sum_{\ell \in \mathcal{L}} {x_\ell \widetilde{\alpha}_\ell}   \\
  & \,\textnormal{subject to} & & \sqrt{ \widetilde{\alpha}_0^2 + \sum_{\ell \in \mathcal{L}} {\widetilde{\alpha}_\ell^2} } \leq \delta  .
\end{alignat}
\end{subequations}
Then, $g(\mathbf{x};\delta) = \delta \sqrt{1 + \sum_{\ell \in \mathcal{L}} {x_\ell}} \,$, which is achieved for $\widetilde{\alpha}_0^\star = \widetilde{\alpha}_\ell^\star = \delta \big/ \sqrt{1 + \sum_{l \in \mathcal{L}} {x_l}} \,$,  $\forall \ell \in \mathcal{P}$, and $\widetilde{\alpha}_\ell^\star = 0$,  $\forall \ell \in {\mathcal{Z}}$, where $\mathcal{P} = \{\ell \in \mathcal{L} : \, x_\ell = 1 \}$ and $\mathcal{Z} = \mathcal{L} \setminus \mathcal{P} = \{\ell \in \mathcal{L} : \, x_\ell = 0 \}$. 
\end{lemma}

\begin{IEEEproof}
Obviously, $|\mathcal{P}| = \sum_{\ell \in \mathcal{L}} {x_\ell}$ and $\sum_{\ell \in \mathcal{L}} {x_\ell \widetilde{\alpha}_\ell} = \sum_{\ell \in \mathcal{P}} {\widetilde{\alpha}_\ell}$. Now, we examine the following relaxation problem 
\begin{subequations} \label{problem:g_rel}
\begin{alignat}{3}
 g_\text{rel}(\mathbf{x};\delta) \! \triangleq & \mathop {\text{maximize}} \limits_{\widetilde{\boldsymbol{\alpha}}_{\mathcal{P}_0} \in \mathbb{R}_{+}^{|\mathcal{P}|+1}} & \quad & \widetilde{\alpha}_0 + \sum_{\ell \in \mathcal{P}} {\widetilde{\alpha}_\ell}   \\ 
  & \,\text{subject to} & & \sqrt{ \widetilde{\alpha}_0^2 + \sum_{\ell \in \mathcal{P}} {\widetilde{\alpha}_\ell^2} } \leq \delta ,
\end{alignat}
\end{subequations}
where $\widetilde{\boldsymbol{\alpha}}_{\mathcal{P}_0} = [{\widetilde{\alpha}_\ell}]_{\ell \in \mathcal{P}_0}^\top$, with ${\mathcal{P}_0} = \{0\} \cup \mathcal{P}$. Since $\sqrt{ \widetilde{\alpha}_0^2 + \sum_{\ell \in \mathcal{P}} {\widetilde{\alpha}_\ell^2} } \leq \sqrt{ \widetilde{\alpha}_0^2 + \sum_{\ell \in \mathcal{L}} {\widetilde{\alpha}_\ell^2} }$, we have that $g(\mathbf{x};\delta) \leq g_\text{rel}(\mathbf{x};\delta)$. On the other hand, an optimal solution of the relaxation problem \eqref{problem:g_rel} together with $\widetilde{\alpha}_\ell = 0$, $\forall \ell \in {\mathcal{Z}}$, constitute a feasible solution of problem \eqref{problem:g}, so $g(\mathbf{x};\delta) \geq g_\text{rel}(\mathbf{x};\delta)$. By combining both results, we obtain $g(\mathbf{x};\delta) = g_\text{rel}(\mathbf{x};\delta)$. 

Subsequently, in order to solve problem \eqref{problem:g_rel}, we make use of the \emph{Cauchy-Bunyakovsky-Schwarz  inequality}: For any $\mathbf{y},\mathbf{z} \in \mathbb{R}^N$, it holds that
\begin{equation}
\left| \sum_{n=1}^N {y_n z_n} \right| \leq \sqrt{\sum_{n=1}^N {y_n^2}} \sqrt{\sum_{n=1}^N {z_n^2}} \, .
\end{equation} 
The above inequality holds with equality if and only if (iff) $\mathbf{y} = \lambda \mathbf{z}$ for some $\lambda \in \mathbb{R}$. Specifically, by setting $\mathbf{z} = \mathbf{1}_N$, we obtain $\left| \sum_{n=1}^N {y_n} \right| \leq \sqrt{N} \sqrt{\sum_{n=1}^N {y_n^2}}$ with equality iff $y_1=\cdots=y_N=\lambda$. Consequently, we obtain  
\begin{equation}
\widetilde{\alpha}_0 + \sum_{\ell \in \mathcal{P}} {\widetilde{\alpha}_\ell} \leq \sqrt{1 + |\mathcal{P}|} \sqrt{ \widetilde{\alpha}_0^2 + \sum_{\ell \in \mathcal{P}} {\widetilde{\alpha}_\ell^2} } \leq \delta \sqrt{1 + |\mathcal{P}|} \, ,
\end{equation}
with equalities (i.e., achieving the maximum value) when $\widetilde{\alpha}_0 = \widetilde{\alpha}_\ell = \lambda \geq 0$, $\forall \ell \in \mathcal{P}$, where $\sqrt{\lambda^2 (1+|\mathcal{P}|)} = \delta \implies \lambda = \delta \big/ \sqrt{1+|\mathcal{P}|}$. Thus, $g(\mathbf{x};\delta) = g_\text{rel}(\mathbf{x};\delta) = \delta \sqrt{1 + |\mathcal{P}|}$.
\end{IEEEproof}

Next, by leveraging Assumption \ref{assumption:minimum_quantization_bits}, the phase-shift quantization error satisfies: i) $|\varepsilon_\ell| \leq \omega/2 \leq \pi/4 \implies \cos(\varepsilon_\ell) \geq \cos(\pi/4) = 1/\sqrt{2} \geq 0$, and ii) from the triangle inequality, $|\varepsilon_n - \varepsilon_m| \leq |\varepsilon_n| + |\varepsilon_m| \leq \omega \leq \pi/2 \implies \cos(\varepsilon_n - \varepsilon_m) \geq 0$. Therefore, $\zeta_\ell = \widehat{\alpha}_\ell^2 + 2 \widehat{\alpha}_0 \widehat{\alpha}_\ell \cos(\varepsilon_\ell) \geq \widehat{\alpha}_\ell^2$ and $\mu_{n m} = 2 \widehat{\alpha}_n \widehat{\alpha}_m \cos(\varepsilon_n - \varepsilon_m) \geq 0$ in \eqref{equation:f_d_sqrt_expanded}. By combining these results with Assumption \ref{assumption:CSI-uncertainty_radius}, we can easily prove that  
\begin{equation} 
\begin{split}
f_\text{d}(\mathbf{x}) & \geq \sqrt{ \widehat{\alpha}_0^2 + \sum_{\ell \in \mathcal{L}}{{\widehat{\alpha}_\ell^2} {x_\ell}} } \geq  \widehat{\alpha}_{\min} \sqrt{1 + \sum_{\ell \in \mathcal{L}} {x_\ell}}  \\
& \geq \delta \sqrt{1 + \sum_{\ell \in \mathcal{L}} {x_\ell}} = g(\mathbf{x};\delta), \;\; \forall \mathbf{x} \in \{0,1\}^L   . 
\end{split} 
\end{equation}
As a result, $f_\text{d}(\mathbf{x}) - \left( \widetilde{\alpha}_0 + \sum_{\ell \in \mathcal{L}} {x_\ell \widetilde{\alpha}_\ell} \right) \geq f_\text{d}(\mathbf{x}) - g(\mathbf{x};\delta) \geq 0$, $\forall \mathbf{x} \in \{0,1\}^L$, $\forall \widetilde{\boldsymbol{\alpha}} \in \mathcal{U}_{\delta}$, where $\mathcal{U}_{\delta} = \left\{ \widetilde{\boldsymbol{\alpha}} \in \mathbb{R}_{+}^{L+1} : \, \sqrt{ \sum_{\ell \in \mathcal{L}_0} {\widetilde{\alpha}_\ell^2} } \leq \delta \right\}$. Therefore, problems \eqref{problem:Absolute_value} and \eqref{problem:g} are equivalent, and the worst-case SNR is given by \eqref{equation:Worst-case_SNR_discrete}. This completes the proof of Theorem \ref{theorem:Worst-case_SNR_discrete}.

\section{Proof of Theorem \ref{theorem:DP_algorithm}}  \label{appendix:DP_algorithm}

In order to prove the correctness of the algorithm, we will use a \emph{loop invariant} \cite{Cormen2009}, namely, 

\vspace{0.7mm}
\noindent $I(M)$: \emph{``At the end of iteration $M$ of the for-loop in steps 9--15, ${\operatorname{EE}^\star} = \operatorname{EE}^\star_{\leq M}$. In addition, if subproblem \eqref{subproblem:less_equal_M} is feasible, then an optimal solution to this subproblem is given by: $x_{\zeta_m}^\star = 1$, $\forall m \in {\mathcal{M}^\star}=\{1,\dots,M^\star\}$, and $x_{\zeta_\ell}^\star = 0$, $\forall \ell \in {\mathcal{L} \setminus {\mathcal{M}^\star}}$.''}
\vspace{0.7mm}

Now, we will show that $I(M)$ is true for all $M \in \mathcal{L}_0$, using mathematical induction. In particular, we can obtain the following: 1) steps 2--8 together with Proposition \ref{proposition:Equivalent_subproblem} imply that $I(0)$ is true before the first iteration of the for-loop (\emph{basis}), and 2) if $I(M-1)$ is true then, by virtue of Proposition \ref{proposition:Equivalent_subproblem} and recurrence relation \eqref{equation:Recurrence_relation}, $I(M)$ is true as well (\emph{inductive step}). Hence, the above argument has been proved.  

After the termination of the for-loop, $I(L)$ will be true. As a result, ${\operatorname{EE}^\star} = \operatorname{EE}^\star_{\leq L}$, but $\operatorname{EE}^\star_{\leq L} = \operatorname{EE}^{\text{c},\star}_\text{worst}$ (see Remark \ref{remark:Recurrence_relation}), thus ${\operatorname{EE}^\star} = \operatorname{EE}^{\text{c},\star}_\text{worst}$. If problem \eqref{problem:EE_c_original} is not feasible (i.e., $\operatorname{EE}^{\text{c},\star}_\text{worst} = -\infty$), then Algorithm \ref{algorithm:DP} correctly returns \textit{``Infeasible''} (steps 16--17). Otherwise, if the problem is feasible (i.e., $\operatorname{EE}^{\text{c},\star}_\text{worst} > -\infty$), then $I(L)$ implies that an optimal solution to problem \eqref{problem:EE_c_original} can be derived using $M^\star$.\footnote{Recall that subproblem \eqref{subproblem:less_equal_M} for $M=L$ coincides with problem \eqref{problem:EE_c_original}, according to Remark \ref{remark:Recurrence_relation}.} This optimal solution is correctly reconstructed and returned by the algorithm (steps 18--24).

Finally, regarding the complexity of Algorithm \ref{algorithm:DP}, the sorting procedure in step 1 requires $O(L\,{\log L})$ comparisons, while all the remaining steps take $O(L)$ time. Consequently, the overall complexity is $O(L\,{\log L} + L) = O(L\,{\log L})$.

\section{Proof of Theorem \ref{theorem:CRBM}} \label{appendix:CRBM}

Firstly, Theorem \ref{theorem:Worst-case_SNR_discrete} is applicable due to Assumptions \ref{assumption:CSI-uncertainty_radius} and \ref{assumption:minimum_quantization_bits}. Secondly, ${\gamma}_{\textnormal{worst}}^\textnormal{d} (\mathbf{1}_L;\delta) \geq \gamma_{\min}$ implies that $\mathbf{x} = \mathbf{1}_L$ is a feasible solution to problem \eqref{problem:EE_d_original}, thus the relaxation problem \eqref{problem:Fractional_relaxation}/\eqref{problem:Convex_relaxation}/\eqref{problem:Convex_relaxation_CVX} is feasible (according to Remark \ref{remark:Fractional_relaxation}).   

Moreover, by construction of Algorithm \ref{algorithm:CRBM}, the returned vector $\widetilde{\mathbf{x}}$ is a feasible solution to problem \eqref{problem:EE_d_original}, since it is binary and satisfies the minimum-SNR constraint, with $\widetilde{\operatorname{EE}} = {\operatorname{EE}_{\textnormal{worst}}^\textnormal{d}}(\widetilde{\mathbf{x}};\delta)$. Note that the programming variable $M_{\max}$ will eventually take some definite value before the for-loop in steps 20-22, due to the assumption that ${\gamma}_{\textnormal{worst}}^\textnormal{d} (\mathbf{1}_L;\delta) \geq \gamma_{\min}$. As a result, the termination of the algorithm is ensured.  

The performance guarantee \eqref{equation:performance_guarantee} can be easily derived from the fact that ${\operatorname{EE}_{\textnormal{worst}}^{\textnormal{d},\star}} \leq \operatorname{EE}_{\textnormal{rel}}^{\star}$ (see Remark \ref{remark:Fractional_relaxation}). Furthermore, the complexity of solving the convex relaxation problem \eqref{problem:Convex_relaxation} using an interior-point method is $O(L^{3.5})$, because this problem has $L+1 = \Theta(L)$ variables and $2L+2 = \Theta(L)$ constraints \cite{Boyd2004}. Observe that problem \eqref{problem:Convex_relaxation_CVX} has the same asymptotic complexity as problem \eqref{problem:Convex_relaxation}, since the number of its variables and constraints are $L+2 = \Theta(L)$ and $2L+3 = \Theta(L)$, respectively. Also, the sorting procedure in step 3 requires $O(L\log{L})$ time, while the remaining steps consist of $O(L)$ arithmetic operations/comparisons. Therefore, the overall complexity of Algorithm \ref{algorithm:CRBM} is $O(L^{3.5} + L\log{L} + L ) = O(L^{3.5})$.

\end{document}